\documentclass[aps,prl,reprint,superscriptaddress]{revtex4-2}
\usepackage{amsmath, amsthm, amssymb}
\usepackage{braket}
\usepackage{dsfont}
\usepackage{mathtools}
\usepackage{amsfonts}
\usepackage{bm}
\usepackage{physics}
\usepackage{comment}
\usepackage{tikz}
\usetikzlibrary{calc}
\usetikzlibrary{math}
\usepackage{ulem}
\usepackage{hyperref}

\newcommand{\rra}{\right\rangle}
\newcommand{\lla}{\left\langle}

\newcommand{\mc}[1]{\mathcal{#1}}

\newcommand{\1}{\mathds{1}}

\newcommand{\Ham}{\mc{H}}

\newcommand\rmi[2]{\ensuremath{#1_{\mathrm{#2}}}}
\newcommand\Fig[1]{Fig.~\ref{#1}}
\newcommand\Eq[1]{Eq.~\eqref{#1}}

\newcommand\Tab[1]{Tab.~\ref{#1}}
\newcommand\Sec[1]{Sec.~\ref{#1}}

\newcommand\sz{\ensuremath{\sigma^z}}
\newcommand\sx{\ensuremath{\sigma^x}}
\newcommand\compbas{\{\ket{\alpha}\}}
\DeclareMathOperator*{\SumInt}{%
\mathchoice%
  {\ooalign{$\displaystyle\sum$\cr\hidewidth$\displaystyle\int$\hidewidth\cr}}
  {\ooalign{\raisebox{.14\height}{\scalebox{.7}{$\textstyle\sum$}}\cr\hidewidth$\textstyle\int$\hidewidth\cr}}
  {\ooalign{\raisebox{.2\height}{\scalebox{.6}{$\scriptstyle\sum$}}\cr$\scriptstyle\int$\cr}}
  {\ooalign{\raisebox{.2\height}{\scalebox{.6}{$\scriptstyle\sum$}}\cr$\scriptstyle\int$\cr}}
}
\newcommand\duc{\rmi{d}{uc}}

\newcommand\wup{\rmi{w}{up}}
\newcommand\wdown{\rmi{w}{down}}
\newcommand\wAF{\rmi{w}{AF}}

\newcommand\Wod{\rmi{W}{od}}
\newcommand\bup{\rmi{b}{\bullet\bullet}}

\newcommand\bod{\rmi{b}{od}}
\newcommand\Bod{\rmi{B}{od}}

\newcommand\ie{i.\,e.\,}
\newcommand\eg{e.\,g.\,}
\newcommand\etal{et\ al.\,}
\usepackage{scalerel,stackengine}

\usepackage[greek,english]{babel}
\newcommand\koppa{\text{\fontfamily{mak}\selectfont \selectlanguage{greek}{\qoppa}}}

\def\uul#1{\vphantom{{\ul{#1}}}\smash{\mathord{\vtop{\ialign{##\crcr
$\sum_{\{|\alpha\rangle\}}\hfil #1\hfil$\crcr\noalign{\kern1.5pt\nointerlineskip}
$\hfil\uline{\hphantom{#1}}\hfil$\crcr\noalign{\kern1.5pt}}}}}}

\definecolor{phase1}{rgb}{0.4795847750865052, 0.7984621299500193, 0.7695501730103806}
\definecolor{phase2}{rgb}{0.3411764705882353, 0.7215686274509804, 0.8156862745098039}
\definecolor{phase3}{rgb}{0.16539792387543253, 0.5456978085351788, 0.7434371395617071}
\definecolor{phase4}{rgb}{0.03137254901960784, 0.3740099961553249, 0.6381391772395233}

\begin{document}

\title{Quantum phase diagrams of Dicke-Ising models by a wormhole algorithm}

\author{Anja Langheld}
\email{anja.langheld@fau.de}
\author{Max Hörmann}
\author{Kai Phillip Schmidt}
\affiliation{Department of Physics, Staudtstra{\ss}e 7, D-91058 Erlangen, Friedrich-Alexander-Universit\"at Erlangen-N\"urnberg (FAU), Germany}

\date{\today}

\begin{abstract}
We gain quantitative insights on effects of light-matter interactions on correlated quantum matter by quantum Monte Carlo simulations. 
We introduce a wormhole algorithm for the paradigmatic Dicke-Ising model which combines the light-matter interaction of the Dicke model with Ising interactions.
The quantum phase diagrams for ferro- and antiferromagnetic interactions on the chain and the square lattice are determined. 
The occurring superradiant phase transitions are in the same universality class as the Dicke model leading to a well-known peculiar finite-size scaling that we elucidate in terms of scaling above the upper critical dimension.
For the ferromagnetic case, the transition between the normal and the superradiant phase is of second order with Dicke criticality (first order) for large (small) longitudinal fields separated by a multicritical point. 
For antiferromagnetic interactions, we establish the light-matter analogue of a lattice supersolid with off-diagonal superradiant and diagonal magnetic order and determine the nature of all transition lines. 
Our achievements allow to quantitatively enter the domain of mesoscopic light-matter systems, which is an increasingly prominent topic at the interface of quantum optics and condensed matter physics.
\end{abstract}

%\keywords{}

\maketitle


The Dicke model \cite{Dicke1954} continues to play a central role in quantum optics as a paradigmatic model for interaction between light and matter displaying a second-order phase transition to a photon condensate with extensive photon number $\langle a^\dagger a\rangle\sim N$ \cite{Hepp1973a, Wang1973, Hepp1973b}.
While in the Dicke model, the matter entities are solely interacting with the bosonic field, from a condensed matter perspective, interactions are an indispensable ingredient to generate highly-correlated quantum states.
With a plethora of exciting quantum many-body phenomena at hand, the question arises 
how the coupling to light elicits new facets of these correlated many-body systems \cite{Schlawin2019, Schuler2020, Paravicini2019, Dmytruk2022, Dmytruk2024}.
This could include the stabilization of exotic quantum phases like quantum spin liquids with topological order or the realization of highly entangled hybrid light-matter states \cite{passetti2023, roman2024linear}.
Inevitably, the interest in the interplay of light-matter and matter-matter interaction has tremendously increased \cite{roman2024linear, Mazza2019, LWeber2023, Schellenberger2024, Schneider2024} and recent progress has been made by mapping a class of light-matter systems to sole matter problem with additional self-consistent constraints in the thermodynamic limit \cite{lenk2022collective,roman2024linear,roman2024cavity}. 
Within this field of correlated light-matter assemblies, the Dicke-Ising model plays a pivotal role due to its paradigmatic nature and experimental relevance \cite{Lee2004, Schellenberger2024, Zhang2014, Schneider2024, Rohn2020, Schuler2020,Gelhausen2016, Zhang2013, Puel2024}, \eg, for Rydberg atoms in an optical cavity \cite{Zhang2013, Puel2024}.

Prominent experimental platforms to realize such light-matter quantum many-body systems are cavity and circuit quantum electrodynamics (QED).
While in atomic cavity QED the existence of systems displaying superradiant phase transitions is often doubted \cite{Rzazewski1975, Nataf2010b, Andolina2019, Andolina2022, Garziano2020},
circuit QED offers an additional extensive playground for the potential realization of such systems \cite{Blais2004, Wallraff2004, You2006, Chen2007, Nataf2010a, Nataf2010b, Bamba2016, Zhang2014, Jaako2016}.
Although certain types of circuits have been shown to also suffer from no-go theorems similar to cavity QED \cite{Viehmann2011}, superradiant phase transitions within circuit QED remain subject to intense discussion \cite{Nataf2010a, Nataf2010b, Viehmann2011, Bamba2016, Andolina2019}.
While in cavity setups the interactions of the two-level entities might be intrinsic, e.g., an Ising interaction in Rydberg atom arrays \cite{Zhang2013,Puel2024}, the artificial atoms in a superconducting circuit can be explicitly coupled.
Ising couplings have been proposed and implemented by circuits of superconducting qubits \cite{Pashkin2003, Zhang2014} realizing a Dicke-Ising model when coupled to a common resonator \cite{Zhang2014}.

Approximate treatments of the Dicke-Ising model indicate a rich quantum phase diagram for antiferromagnetic Ising interactions including 
the presence of an intermediate phase with simultaneous off-diagonal superradiant and diagonal magnetic order, which can be seen as a light-matter analogue of a lattice supersolid \cite{Zhang2014, Zhang2013, Gelhausen2016}. 
However, a quantitative description of this quantum many-body system coupled to a resonator, in particular beyond 1d, has not been achieved. 
The quantitative treatment of finite mesoscopic Dicke-like systems poses a tremendous challenge and even for the Dicke model this has been achieved only after several decades \cite{Lambert2004, Vidal2006, Liberti2006, Chen2008}. Adding inherently different matter interactions to the problem makes this even more challenging.
In this letter we present an adaption of the wormhole quantum Monte Carlo algorithm \cite{Weber2017, Weber2022b} that is capable to accurately determine ground-state properties of correlated quantum many-body systems strongly coupled to light. 
This allows us to provide quantitative quantum phase diagrams for the (anti-)ferromagnetic Dicke-Ising model in one and two dimensions.
We characterize the nature of all occurring phase transitions and additionally bring the criticality and finite-size scaling of superradiant phase transitions into line with expectations.

{\it Dicke-Ising model.} 
The Dicke-Ising model (DIM) combines the Dicke model of $N$ spins-1/2 coupled to a common bosonic light mode with a nearest-neighbor Ising interaction among the spins. It is given by
\begin{align}
	\Ham =  \omega a^\dagger a  + \epsilon \sum_{i=1}^N{\sz_i} + \frac{g}{\sqrt{N}} (a^\dagger + a) \sum_{i=1}^N \sigma^x_i + J \sum_{\lla i, j\rra} \sz_i \sz_j
	\label{eq:dicke-ising}
\end{align}
with $\sigma_i^{x/z}$ the Pauli matrices describing spins and $a^{(\dagger)}$ the bosonic annihilation (creation) operator describing photons of frequency $\omega$. 
The system's parameters are tuned by the longitudinal field $\epsilon$, the spin-photon coupling $g$, and the Ising coupling $J$, which can either be ferromagnetic $J<0$ or antiferromagnetic $J>0$.
Note that one can easily include the effect of a diamagnetic $A^2$-term by absorbing the term into the photon frequency $\omega$ \cite{FriskKockum2019} (see \cite{sm} for adding this term post simulation).

For $\epsilon = 0$, the DIM possesses a $\mathbb{Z}_2 \cross \mathbb{Z}_2$ symmetry.
One of the $\mathbb{Z}_2$ symmetries is the spin-flip symmetry $\sz \rightarrow -\sz$ inherent to the Ising model and the other one is the parity $(a, \sx) \rightarrow (-a, -\sx)$ inherent to the Dicke model. 
For $\epsilon > 0$, the spin-flip symmetry $\sz \rightarrow -\sz$ of the Ising model is not present anymore and the ferromagnetically ordered state does not break any symmetry of the Hamiltonian.
In contrast, for antiferromagnetic Ising interactions, the staggered spin state also breaks the translational symmetry of the lattice for finite $\epsilon$.

The zero-temperature properties of the DIM have been studied in one dimension for $\epsilon = 0$ \cite{Rohn2020}, within mean field theory \cite{Zhang2014, Puel2024}, and by quantum Monte Carlo (QMC) simulations within the rotating-wave approximation \cite{Zhang2013}.
On the linear chain and $\epsilon=0$, the DIM was shown to undergo a first-order phase transition between a magnetically ordered phase with vanishing photon density and a paramagnetic superradiant phase with finite photon density \cite{Rohn2020,roman2024linear}.
The first-order nature of this phase transition between two ordered phases with different broken symmetries is expected by Landau theory.
For ferromagnetic Ising interactions in the presence of a longitudinal field, the phase diagram was determined within a self-consistent mean field approach claiming that the transition is of first order \cite{Puel2024}.
The phase diagram of the antiferromagnetic model for general longitudinal field was studied using a variational mean field ansatz decoupling light and matter as well as treating the matter degrees of freedom as classical spins \cite{Zhang2014}. 
In addition to the antiferromagnetic normal phase and paramagnetic superradiant phase already present at $\epsilon =0$, one finds a completely disordered paramagnetic normal phase and an intermediate antiferromagnetic superradiant phase breaking both symmetries. 
This latter phase has diagonal and off-diagonal order simultaneously and can be seen as the light-matter analogue of lattice supersolids known from frustrated quantum magnets \cite{Momoi2000,Kwai-Kong2006,Sengupta2007, Laflorencie2007, KPS2008} as well as trapped cold atoms \cite{Wessel2005, Heidarian2005, Melko2005}. 
We stress that, in contrast to the antiferromagnetic superradiant phase, the conventional lattice supersolid breaks a continuous U(1) symmetry and therefore has gapless excitations. Interestingly, in \cite{Zhang2014} all quantum phase transitions are predicted to be of second order, which is in conflict with the quantitative investigation in one dimension for $\epsilon=0$ \cite{Rohn2020}.  
A proper investigation of the full DIM for arbitrary Ising interactions is therefore needed.

{\it Wormhole quantum Monte Carlo.} To this end we tailored the QMC algorithm based on Refs.~\cite{Weber2017, Weber2022b} for spin-boson models to the DIM.
The key idea is to first integrate out the bosons and only simulate the resulting effective spin model \cite{Weber2017, Weber2022b}.
For the DIM, one obtains
\begin{equation}
	\label{eq:ret_action}
	\begin{aligned}
	\rmi{S}{ret} &= \int_0^\beta \dd{\tau} \left[ \epsilon \sum_{i=1}^N \sz_i(\tau) + J\sum_{\langle i,j \rangle} \sz_i(\tau)\sz_j(\tau)\right] -\\
	&\frac{g^2}{\omega N} \int_0^\beta \dd{\tau}\int_0^\beta \dd{\tau'} \sum_{i,j=1}^N \sx_i(\tau') P(\omega, \tau' - \tau) \sx_j (\tau)
	\end{aligned}
\end{equation}
with the (normalized) free boson propagator $P(\omega, \Delta\tau) = \omega \frac{e^{-\Delta\tau \omega}}{1- e^{-\beta\omega}}$.
The partition function $Z = Z_{\rm b} \Tr_{\rm s}\left[ \mathcal{T}_{\tau} e^{-\rmi{S}{ret}} \right]$ with $Z_{\rm b}$ the partition function of free bosons and spin trace $\Tr_{\rm s}\left[\dots\right]$ is then rewritten as in the stochastic series expansion QMC \cite{Sandvik1991, Sandvik2010} 
			\begin{align}
				\frac{Z}{Z_{\rm b}} &= \sum_{\compbas} \sum_{n=0}^{\infty} \sum_{\{C_n\}}\frac{1}{n!} \bra{\alpha} \mathcal{T}_\tau  \prod_{k=1}^n \rmi{\Ham}{\nu_k}\dd{\tau_k}\dd{\tau'_k}\ket{\alpha}
			\end{align}
with a suitable decomposition of $\rmi{S}{ret} =  -\SumInt_{\nu} \Ham_\nu\dd{\tau}\dd{\tau'}$ into operators $\Ham_\nu$ with vertex variables $\nu = \{\mathrm{type}, i,j, \tau, \tau'\}$ and the ordered vertex list $C_n = \{ \nu_1, \dots , \nu_n\}$ \cite{Weber2022b}. 
The computational basis $\compbas$ is chosen to be the Ising basis $\ket{\alpha} = \bigotimes_{i=1}^N \ket{\sz}_i$ and $\rmi{S}{ret}$ is decomposed into diagonal (d) and off-diagonal (od) parts acting on pairs of spins $i,j$
\begin{align}
	\Ham^{\mathrm{d}}_{ij}(\tau, \tau') = & \frac{\delta_{\tau, \tau'}}{2} 
	\begin{aligned}[t]
		\left[ \vphantom{\frac{\epsilon}{z}}\right.&\abs{J}-J\sz_i(\tau) \sz_j (\tau) + \\
		&\left.\frac{\epsilon}{z}(2- \sz_i(\tau) - \sz_j(\tau))\right]
	\end{aligned}\\
	\Ham^{\mathrm{od}}_{ij}(\tau, \tau') =& \frac{g^2}{\omega N} \sx_i(\tau') P(\omega, \tau' - \tau) \sx_j (\tau).
\end{align}
%auto-ignore

\def\radius{2pt}
\def\lw{1pt}

\def\emptycircle{\tikz[baseline=-0.5ex]{
	\draw [line width = \lw] (0,0.2ex) circle (\radius);
}}
\def\fullcircle{\tikz[baseline=-.5ex]{
	\fill (0,0.2ex) circle (\radius);
}}

\def\diagop(#1,#2,#3,#4){
	\def\x{#1}
	\def\y{#2}
	\def\dx{#3}
	\def\dy{#4}
	\draw [line width = \lw] (\x + \radius, \y) circle (\radius) coordinate (A);
	\draw [line width = \lw] (\x + \radius, \y + \dy) circle (\radius) coordinate (B);
	\draw [line width = \lw] (\x + \dx + \radius, \y + \dy) circle (\radius) coordinate (C);
	\draw [line width = \lw] (\x + \dx + \radius, \y) circle (\radius) coordinate (D);
	\fill (\x -.5pt, \y + 0.5*\dy - .5pt) rectangle ++(\dx + 5pt, 1pt);
}

\def\offdiagop(#1,#2,#3,#4){
	\def\xl{#1}
	\def\xr{#2}
	\def\yu{#3}
	\def\yl{#4}
	\draw [line width = \lw] (\xl + \radius, \yl - 2.5*\radius) circle (\radius) coordinate (A);
	\draw [line width = \lw] (\xl + \radius, \yl + 2.5*\radius) circle (\radius) coordinate (B);
	\draw [line width = \lw] (\xr + \radius, \yu + 2.5*\radius) circle (\radius) coordinate (C);
	\draw [line width = \lw] (\xr + \radius, \yu - 2.5*\radius) circle (\radius) coordinate (D);
	\fill (\xl -.5pt,\yl - .5pt) rectangle ++(5pt, 1pt);
	\fill (\xr -.5pt,\yu - .5pt) rectangle ++(5pt, 1pt);
	\draw [dashed, line width = \lw] ($(A) + (\radius, 2.5*\radius)$) -- ($(D) + (-\radius, 2.5*\radius)$);
}

\def\labelsites{
	\node[anchor = north] at (\x + \radius, \y - 5*\radius) {$i$};
	\node[anchor = north] at (\x + \dx + \radius, \y - 5*\radius) {$j$};
}
\def\labelodsites{
	\node[anchor = north] at (\xl + \radius, \yl - 6*\radius) {$i$};
	\node[anchor = north] at (\xr + \radius, \yl - 6*\radius) {$k$};
}
\def\labelimagtime{
	\node[anchor = east] at ($0.5*(B) + 0.5*(A) - (1ex,  0)$) {$\tau$};
	\node[anchor = west] at ($0.5*(D) + 0.5*(C) + (1ex, 0) $) {$\tau+\Delta \tau$};
}

\def\drawlegs{
	\draw [line width = \lw] (\x + \radius, \y - 5*\radius) -- ($(A) - (0, 4pt)$);
	\draw [line width = \lw] (\x + \radius, \y + \dy + 5*\radius) -- ($(B) + (0, 4pt)$);
	\draw [line width = \lw] (\x + \dx + \radius, \y + \dy + 5*\radius) -- ($(C) + (0, 4pt)$);
	\draw [line width = \lw] (\x + \dx + \radius, \y - 5*\radius) -- ($(D) - (0, 4pt)$);
}
\def\drawodlegs{
	\draw [line width = \lw] (\xl + \radius, \yl - 6*\radius) -- ($(A) - (0, 4pt)$);
	\draw [line width = \lw] (\xl + \radius, \yu + 6*\radius) -- ($(B) + (0, 4pt)$);
	\draw [line width = \lw] (\xr + \radius, \yu + 6*\radius) -- ($(C) + (0, 4pt)$);
	\draw [line width = \lw] (\xr + \radius, \yl - 6*\radius) -- ($(D) - (0, 4pt)$);
}

\def\diagopwlegs(#1,#2,#3,#4){
	\diagop(#1,#2,#3,#4)
	\drawlegs
}

\def\offdiagopwlegs(#1,#2,#3,#4){
	\offdiagop(#1,#2,#3,#4)
	\drawodlegs
	%\labelodsites
}

\def\Diagopd{\tikz[baseline = 0, scale = 0.6]{
	\def\lw{0.7pt}
	\diagop(0pt,0pt,10pt, 10pt)
	\fill (\x + \radius, \y) circle (\radius) coordinate (A);
	\fill (\x + \radius, \y + \dy) circle (\radius) coordinate (B);
	\fill (\x + \dx + \radius, \y + \dy) circle (\radius) coordinate (C);
	\fill (\x + \dx + \radius, \y) circle (\radius) coordinate (D);
}}
\def\Diagopaf{\tikz[baseline = 0, scale = 0.7]{
	\def\lw{0.7pt}
	\diagop(0pt,0pt,10pt, 10pt)
	\fill (\x + \radius, \y) circle (\radius) coordinate (A);
	\fill (\x + \radius, \y + \dy) circle (\radius) coordinate (B);
}}
\def\Diagopafr{\tikz[baseline = 0, scale = 0.7]{
	\def\lw{0.7pt}
	\diagop(0pt,0pt,10pt, 10pt)
	\fill (\x + \dx + \radius, \y + \dy) circle (\radius) coordinate (C);
	\fill (\x + \dx + \radius, \y) circle (\radius) coordinate (D);
}}
\def\Diagopu{\tikz[baseline = 0, scale = 0.7]{
	\def\lw{0.7pt}
	\diagop(0pt,0pt,10pt, 10pt)
}}
\def\Offdiagop{\tikz[baseline = 0, scale = 0.7]{
	\def\lw{0.7pt}
	\offdiagop(00pt,15pt,10pt,01pt)
	\fill (B) circle (\radius);
	\fill ($(C) + (\radius/1.41,\radius/1.41)$) arc (45:225:\radius) -- cycle;
	\fill ($(D) - (\radius/1.41,\radius/1.41)$) arc (225:405:\radius) -- cycle;
}}
\def\Offdiagopu{\tikz[baseline = 0, scale = 0.7]{
	\def\lw{0.7pt}
	\offdiagop(00pt,15pt,10pt,01pt)
	\fill (B) circle (\radius);
	\fill ($(D) + (\radius/1.41,\radius/1.41)$) arc (45:225:\radius) -- cycle;
	\fill ($(C) - (\radius/1.41,\radius/1.41)$) arc (225:405:\radius) -- cycle;
}}

\def\diagopu(#1,#2){
	\diagopwlegs(#1,#2,10pt, 10pt)
}
\def\diagopaf(#1,#2){
	\diagopwlegs(#1,#2,10pt, 10pt)
	\fill (\x + \radius, \y) circle (\radius) coordinate (A);
	\fill (\x + \radius, \y + \dy) circle (\radius) coordinate (B);
}
\def\diagopafr(#1,#2){
	\diagopwlegs(#1,#2,10pt, 10pt)
	\fill (\x + \dx + \radius, \y) circle (\radius) coordinate (C);
	\fill (\x + \dx + \radius, \y + \dy) circle (\radius) coordinate (D);
}

\def\diagopd(#1,#2){
	\diagopwlegs(#1,#2,10pt, 10pt)
	\fill (\x + \radius, \y) circle (\radius) coordinate (A);
	\fill (\x + \radius, \y + \dy) circle (\radius) coordinate (B);
	\fill (\x + \dx + \radius, \y + \dy) circle (\radius) coordinate (C);
	\fill (\x + \dx + \radius, \y) circle (\radius) coordinate (D);
}

\def\offdiagnorm(#1,#2){
	\diagopwlegs(#1,#2,10pt, 10pt)
	%\fill (\x + \radius, \y) circle (\radius) coordinate (A);
	\fill (\x + \radius, \y + \dy) circle (\radius) coordinate (B);
	\draw (\x + \dx + \radius, \y + \dy) circle (\radius) coordinate (C);
	\fill (\x + \dx + \radius, \y) circle (\radius) coordinate (D);
}
\def\offdiagnormwocons(#1,#2){
	\diagopwlegs(#1,#2,10pt, 10pt)
	%\fill (\x + \radius, \y) circle (\radius) coordinate (A);
	\fill (\x + \radius, \y + \dy) circle (\radius) coordinate (B);
	\fill (\x + \dx + \radius, \y + \dy) circle (\radius) coordinate (C);
	\draw (\x + \dx + \radius, \y) circle (\radius) coordinate (D);
}

\def\offdiagophalf(#1,#2,#3,#4){
	\offdiagopwlegs(#1,#2,#3,#4)
	\labelimagtime
	\fill ($(D) + (\radius/1.41,\radius/1.41)$) arc (45:225:\radius) -- cycle;
	\fill ($(C) - (\radius/1.41,\radius/1.41)$) arc (225:405:\radius) -- cycle;
	\fill ($(A) + (-\radius/1.41,\radius/1.41)$) arc (135:315:\radius) -- cycle;
	\fill ($(B) - (-\radius/1.41,\radius/1.41)$) arc (315:495:\radius) -- cycle;
}

%%% Worm processes %%%
\def\Offdiagnorm{\tikz[baseline=.5ex+.5pt]{
	\offdiagnorm(0pt,0pt)
	\labelsites
	%\draw [red, -stealth, line width = \lw] ($(B) + (-2*\radius, 10pt)$) -- ($(B) + (-2*\radius, 0pt)$);
	%\draw [red, -stealth, line width = \lw] ($(A) + (-2*\radius, 0pt)$) -- ($(A) + (-2*\radius, -10pt)$);
	\draw [red, -stealth, line width = \lw] ($(C) + (2*\radius, 0pt)$) -- ($(C) + (2*\radius, 10pt)$);
	\draw [red, -stealth, line width = \lw] ($(A) + (-2*\radius, -10pt)$) -- ($(A) + (-2*\radius, 0pt)$);
}}
\def\Offdiagturn{\tikz[baseline=.5ex+.5pt]{
	\offdiagnormwocons(0pt,0pt)
	\labelsites
	%\draw [red, -stealth, line width = \lw] ($(B) + (-2*\radius, 10pt)$) -- ($(B) + (-2*\radius, 0pt)$);
	%\draw [red, -stealth, line width = \lw] ($(A) + (-2*\radius, 0pt)$) -- ($(A) + (-2*\radius, -10pt)$);
	\draw [red, -stealth, line width = \lw] ($(D) + (2*\radius, 0pt)$) -- ($(D) + (2*\radius, -10pt)$);
	\draw [red, -stealth, line width = \lw] ($(A) + (-2*\radius, -10pt)$) -- ($(A) + (-2*\radius, 0pt)$);
}}
\def\Ding{\tikz[baseline=.5ex+.5pt]{
	\diagopd(0pt,0pt)
	\labelsites
	\draw [red, -stealth, line width = \lw] ($(A) - (2*\radius, 10pt)$) -- ($(A) + (-2*\radius, 0pt)$);
}}
\def\Dbounce{\tikz[baseline=.5ex+.5pt]{
	\diagopd(0pt,0pt)
	\labelsites
	\draw [red, -stealth, line width = \lw] ($(A) - (2*\radius, 10pt)$) -- ($(A) + (-2*\radius, 0pt)$);
	\draw [red, -stealth, line width = \lw] ($(A) - (-2*\radius, 0pt)$) -- ($(A) - (-2*\radius, 10pt)$);
}}
\def\Dtraverse{\tikz[baseline=.5ex+.5pt]{
	\diagopd(0pt,0pt)
	\labelsites
	\draw [red, line width = \lw] (A) circle (\radius);
	\draw [red, line width = \lw] (B) circle (\radius);
	\draw [red, -stealth, line width = \lw] ($(B) + (-2*\radius, 10pt)$) -- ($(B) + (-2*\radius, 0pt)$);
	\draw [red, -stealth, line width = \lw] ($(A) + (-2*\radius, 0pt)$) -- ($(A) + (-2*\radius, -10pt)$);
}}

\def\AFingd{\tikz[baseline=.5ex+.5pt]{
	\diagopafr(0pt,0pt)
	\labelsites
	\draw [red, -stealth, line width = \lw] ($(B) + (-2*\radius, 10pt)$) -- ($(B) + (-2*\radius, 0pt)$);
}}
\def\AFbounced{\tikz[baseline=.5ex+.5pt]{
	\diagopafr(0pt,0pt)
	\labelsites
	\draw [red, -stealth, line width = \lw] ($(B) + (-2*\radius, 10pt)$) -- ($(B) + (-2*\radius, 0pt)$);
	\draw [red, -stealth, line width = \lw] ($(B) + (2*\radius, 0pt)$) -- ($(B) + (2*\radius, 10pt)$);
}}
\def\AFrtraverse{\tikz[baseline=.5ex+.5pt]{
	\diagopafr(0pt,0pt)
	\labelsites
	\draw [red, line width = \lw] (A) circle (\radius);
	\draw [red, line width = \lw] (B) circle (\radius);
	\draw [red, -stealth, line width = \lw] ($(B) + (-2*\radius, 0pt)$) -- ($(B) + (-2*\radius, 10pt)$);
	\draw [red, -stealth, line width = \lw] ($(A) + (-2*\radius, -10pt)$) -- ($(A) + (-2*\radius, 0pt)$);
}}

\def\AFing{\tikz[baseline=.5ex+.5pt]{
	\diagopaf(0pt,0pt)
	\labelsites
	\draw [red, -stealth, line width = \lw] ($(A) - (2*\radius, 10pt)$) -- ($(A) - (2*\radius, 0pt)$);
}}
\def\AFbounce{\tikz[baseline=0.5ex+.5pt]{
	\diagopaf(0pt,0pt)
	\labelsites
	\draw [red, -stealth, line width = \lw] ($(A) - (2*\radius, 10pt)$) -- ($(A) - (2*\radius, 0pt)$);
	\draw [red, -stealth, line width = \lw] ($(A) - (-2*\radius, 0pt)$) -- ($(A) - (-2*\radius, 10pt)$);
}}
\def\AFtraverse{\tikz[baseline=.5ex+.5pt]{
	\diagopaf(0pt,0pt)
	\labelsites
	\draw [red, line width = \lw] (A) circle (\radius);
	\draw [red, line width = \lw] (B) circle (\radius);
	\draw [red, -stealth, line width = \lw] ($(B) + (-2*\radius, 10pt)$) -- ($(B) + (-2*\radius, 0pt)$);
	\draw [red, -stealth, line width = \lw] ($(A) + (-2*\radius, 0pt)$) -- ($(A) + (-2*\radius, -10pt)$);
}}

\def\Uingd{\tikz[baseline=.5ex+.5pt]{
	\diagopu(0pt,0pt)
	\labelsites
	\draw [red, -stealth, line width = \lw] ($(B) + (-2*\radius, 10pt)$) -- ($(B) + (-2*\radius, 0pt)$);
}}
\def\Ubounce{\tikz[baseline=.5ex+.5pt]{
	\diagopu(0pt,0pt)
	\labelsites
	\draw [red, -stealth, line width = \lw] ($(B) + (-2*\radius, 10pt)$) -- ($(B) + (-2*\radius, 0pt)$);
	\draw [red, -stealth, line width = \lw] ($(B) + (2*\radius, 0pt)$) -- ($(B) + (2*\radius, 10pt)$);
}}
\def\Utraverse{\tikz[baseline=0.5ex+0.5pt]{
	\diagopu(0pt,0pt)
	\labelsites
	\draw [red, line width = \lw] (A) circle (\radius);
	\draw [red, line width = \lw] (B) circle (\radius);
	\draw [red, -stealth, line width = \lw] ($(A) - (2*\radius, 10pt)$) -- ($(A) - (2*\radius, 0pt)$);
	\draw [red, -stealth, line width = \lw] ($(B) + (-2*\radius, 0pt)$) -- ($(B) + (-2*\radius, 10pt)$);
}}

\def\AFscatterx{\tikz[baseline=.5ex+.5pt]{
	\diagopafr(0pt, 0pt)
	\labelsites
	\fill [red] (\x -.5pt, \y + 0.5*\dy - .5pt) rectangle ++(\dx + 5pt, 1pt);
	\draw [red, line width = \lw] (B) circle (\radius);
	\draw [red, -stealth, line width = \lw] ($(B) - (2*\radius, 0pt)$) -- ($(B) + (-2*\radius, 10pt)$);
}}
\def\Dscatter{\tikz[baseline=.5ex+.5pt]{
	\diagopd(0pt, 0pt)
	\labelsites
	\fill [red] (\x -.5pt, \y + 0.5*\dy - .5pt) rectangle ++(\dx + 5pt, 1pt);
	\draw [red, line width = \lw] (A) circle (\radius);
	\draw [red, -stealth, line width = \lw] ($(A) - (2*\radius, 0pt)$) -- ($(A) - (2*\radius, 10pt)$);
}}
\def\Uscatter{\tikz[baseline=.5ex+.5pt]{
	\diagopu(0pt, 0pt)
	\labelsites
	\fill [red] (\x -.5pt, \y + 0.5*\dy - .5pt) rectangle ++(\dx + 5pt, 1pt);
	\draw [red, line width = \lw] (B) circle (\radius);
	\draw [red, -stealth, line width = \lw] ($(B) - (2*\radius, 0pt)$) -- ($(B) + (-2*\radius, 10pt)$);
}}
\def\AFscattero{\tikz[baseline=.5ex+.5pt]{
	\diagopaf(0pt, 0pt)
	\labelsites
	\fill [red] (\x -.5pt, \y + 0.5*\dy - .5pt) rectangle ++(\dx + 5pt, 1pt);
	\draw [red, line width = \lw] (A) circle (\radius);
	\draw [red, -stealth, line width = \lw] ($(A) + (-2*\radius, 0pt)$) -- ($(A) + (-2*\radius, -10pt)$);
}}

\def\offdiagscatterup{\tikz[baseline=.5ex+.5pt]{
	\offdiagopwlegs(00pt,15pt,10pt,01pt)
	\labelodsites
	\fill (B) circle (\radius);
	\fill ($(C) + (\radius/1.41,\radius/1.41)$) arc (45:225:\radius) -- cycle;
	\fill ($(D) - (\radius/1.41,\radius/1.41)$) arc (225:405:\radius) -- cycle;
	\fill [red, line width = \lw] ($0.5*(A)+0.5*(B) + ( -2.5pt, - .5pt)$) rectangle ++(5pt, 1pt);
	\fill [red, line width = \lw] ($0.5*(C) + 0.5*(D) + ( -2.5pt, - .5pt)$) rectangle ++(5pt, 1pt);
	\draw [red, line width = \lw] (A) circle (\radius);
	\draw [red, line width = \lw] (C) circle (\radius);
	\draw [red, -stealth, line width = \lw] ($(A) - (2*\radius, 10pt)$) -- ($(A) - (2*\radius, 0pt)$);
	\draw [red, -stealth, line width = \lw] ($(C) + (2*\radius, 0pt)$) -- ($(C) + (2*\radius, 10pt)$);
}}
\def\offdiagscatterdown{\tikz[baseline=.5ex+.5pt]{
	\offdiagopwlegs(00pt,15pt,10pt,01pt)
	\labelodsites
	\fill (B) circle (\radius);
	\fill ($(D) + (\radius/1.41,\radius/1.41)$) arc (45:225:\radius) -- cycle;
	\fill ($(C) - (\radius/1.41,\radius/1.41)$) arc (225:405:\radius) -- cycle;
	\fill [red, line width = \lw] ($0.5*(A) + 0.5*(B) + ( -2.5pt, - .5pt)$) rectangle ++(5pt, 1pt);
	\fill [red, line width = \lw] ($0.5*(C) + 0.5*(D) + ( -2.5pt, - .5pt)$) rectangle ++(5pt, 1pt);
	\draw [red, line width = \lw] (A) circle (\radius);
	\draw [red, line width = \lw] (D) circle (\radius);
	\draw [red, -stealth, line width = \lw] ($(A) - (2*\radius, 10pt)$) -- ($(A) - (2*\radius, 0pt)$);
	\draw [red, -stealth, line width = \lw] ($(D) + (2*\radius, 0pt)$) -- ($(D) + (2*\radius, -19pt)$);
}}
\def\offdiagscatterupfromup{\tikz[baseline=.5ex+.5pt]{
	\offdiagopwlegs(00pt,15pt,10pt,01pt)
	\labelodsites
	\fill (B) circle (\radius);
	\fill ($(C) + (\radius/1.41,\radius/1.41)$) arc (45:225:\radius) -- cycle;
	\fill ($(D) - (\radius/1.41,\radius/1.41)$) arc (225:405:\radius) -- cycle;
	\fill [red, line width = \lw] ($0.5*(A) + 0.5*(B) + ( -2.5pt, - .5pt)$) rectangle ++(5pt, 1pt);
	\fill [red, line width = \lw] ($0.5*(C) + 0.5*(D) + ( -2.5pt, - .5pt)$) rectangle ++(5pt, 1pt);
	\draw [red, line width = \lw] (B) circle (\radius);
	\draw [red, line width = \lw] (C) circle (\radius);
	\draw [red, -stealth, line width = \lw] ($(B) + (-2*\radius, 20pt)$) -- ($(B) + (-2*\radius, 0pt)$);
	\draw [red, -stealth, line width = \lw] ($(C) - (-2*\radius, 0pt)$) -- ($(C) + (2*\radius, 10pt)$);
}}

\def\offdiagscatterdownfromup{\tikz[baseline=.5ex+.5pt]{
	\offdiagopwlegs(00pt,15pt,10pt,01pt)
	\labelodsites
	\fill (B) circle (\radius);
	\fill ($(D) + (\radius/1.41,\radius/1.41)$) arc (45:225:\radius) -- cycle;
	\fill ($(C) - (\radius/1.41,\radius/1.41)$) arc (225:405:\radius) -- cycle;
	\fill [red, line width = \lw] ($0.5*(A) + 0.5*(B) + ( -2.5pt, - .5pt)$) rectangle ++(5pt, 1pt);
	\fill [red, line width = \lw] ($0.5*(C) + 0.5*(D) + ( -2.5pt, - .5pt)$) rectangle ++(5pt, 1pt);
	\draw [red, line width = \lw] (B) circle (\radius);
	\draw [red, line width = \lw] (D) circle (\radius);
	\draw [red, -stealth, line width = \lw] ($(B) + (-2*\radius, 20pt)$) -- ($(B) + (-2*\radius, 0pt)$);
	\draw [red, -stealth, line width = \lw] ($(D) - (-2*\radius, 0pt)$) -- ($(D) - (-2*\radius, 20pt)$);
}}

\def\offdiaging{\tikz[baseline=.5ex+.5pt]{
	\offdiagopwlegs(00pt,15pt,10pt,01pt)
	\labelodsites
	\fill (B) circle (\radius);
	\fill ($(C) + (\radius/1.41,\radius/1.41)$) arc (45:225:\radius) -- cycle;
	\fill ($(D) - (\radius/1.41,\radius/1.41)$) arc (225:405:\radius) -- cycle;
	\draw [red, -stealth, line width = \lw] ($(C) + (2*\radius, 10pt)$) -- ($(C) + (2*\radius, 0pt)$) ;
}}
\def\offdiagingup{\tikz[baseline=.5ex+.5pt]{
	\offdiagopwlegs(00pt,15pt,10pt,01pt)
	\labelodsites
	\fill (B) circle (\radius);
	\fill ($(D) + (\radius/1.41,\radius/1.41)$) arc (45:225:\radius) -- cycle;
	\fill ($(C) - (\radius/1.41,\radius/1.41)$) arc (225:405:\radius) -- cycle;
	\draw [red, -stealth, line width = \lw] ($(D) + (2*\radius, -20pt)$) -- ($(D) + (2*\radius, 0pt)$) ;
}}

\def\offdiagbounce{\tikz[baseline=.5ex+.5pt]{
	\offdiagopwlegs(00pt,15pt,10pt,01pt)
	\labelodsites
	\fill (B) circle (\radius);
	\fill ($(C) + (\radius/1.41,\radius/1.41)$) arc (45:225:\radius) -- cycle;
	\fill ($(D) - (\radius/1.41,\radius/1.41)$) arc (225:405:\radius) -- cycle;
	\draw [red, -stealth, line width = \lw] ($(C) + (2*\radius, 10pt)$) -- ($(C) + (2*\radius, 0pt)$) ;
	\draw [red, -stealth, line width = \lw] ($(C) + (-2*\radius, 0pt)$) -- ($(C) + (-2*\radius, 10pt)$) ;
}}

\def\offdiagbounceup{\tikz[baseline=.5ex+.5pt]{
	\offdiagopwlegs(00pt,15pt,10pt,01pt)
	\labelodsites
	\fill (B) circle (\radius);
	\fill ($(D) + (\radius/1.41,\radius/1.41)$) arc (45:225:\radius) -- cycle;
	\fill ($(C) - (\radius/1.41,\radius/1.41)$) arc (225:405:\radius) -- cycle;
	\draw [red, -stealth, line width = \lw] ($(D) + (-2*\radius, 0pt)$) -- ($(D) + (-2*\radius, -20pt)$) ;
	\draw [red, -stealth, line width = \lw] ($(D) + (2*\radius, -20pt)$) -- ($(D) + (2*\radius, 0pt)$) ;
}}
\def\offdiagtraverse{\tikz[baseline=.5ex+.5pt]{
	\offdiagopwlegs(00pt,15pt,10pt,01pt)
	\labelodsites
	\fill (B) circle (\radius);
	\fill ($(D) + (\radius/1.41,\radius/1.41)$) arc (45:225:\radius) -- cycle;
	\fill ($(C) - (\radius/1.41,\radius/1.41)$) arc (225:405:\radius) -- cycle;
	\draw [red, line width = \lw] (C) circle (\radius);
	\draw [red, line width = \lw] (D) circle (\radius);
	\draw [red, -stealth, line width = \lw] ($(C) + (2*\radius, 10pt)$) -- ($(C) + (2*\radius, 0pt)$) ;
	\draw [red, -stealth, line width = \lw] ($(D) + (2*\radius, 0pt)$) -- ($(D) + (2*\radius, -20pt)$) ;
}}
\def\offdiagtraverseup{\tikz[baseline=.5ex+.5pt]{
	\offdiagopwlegs(00pt,15pt,10pt,01pt)
	\labelodsites
	\fill (B) circle (\radius);
	\fill ($(C) + (\radius/1.41,\radius/1.41)$) arc (45:225:\radius) -- cycle;
	\fill ($(D) - (\radius/1.41,\radius/1.41)$) arc (225:405:\radius) -- cycle;
	\draw [red, line width = \lw] (C) circle (\radius);
	\draw [red, line width = \lw] (D) circle (\radius);
	\draw [red, -stealth, line width = \lw] ($(C) + (2*\radius, 0pt)$) -- ($(C) + (2*\radius, 10pt)$) ;
	\draw [red, -stealth, line width = \lw] ($(D) + (2*\radius, -20pt)$) -- ($(D) + (2*\radius, 0pt)$) ;
}}

Exemplary interaction vertices for such operators acting locally on states of the computational basis are visualized in \Fig{fig:operator-vertices}.
\begin{figure}[h]
\begin{tikzpicture}
\draw [-stealth, line width = 0.5*\lw] (-3ex, -12*\radius) coordinate (E) -- (-3ex, 11*\radius) coordinate (F);
\node[anchor = east, align = center] at ($(F) - (1ex, 4*\radius)$) {imag.\\ time};
\draw [-stealth, line width = 0.5*\lw] (-3ex + 2*\radius, -14*\radius) coordinate (E) -- (120pt, -14*\radius) coordinate (F);
\node[anchor = west, align = center] at ($(F) + (1ex, 0)$) {spin sites};
\diagopaf(15pt, 0pt)
\node[anchor = east] at ($0.5*(B) + 0.5*(A) - (1ex,  0)$) {$\tau_1$};
\node[anchor = north] at (\x + \radius, \y - 5*\radius) {$i_1$};
\node[anchor = north] at (\x + \dx + \radius, \y - 5*\radius) {$j_1$};
\offdiagopwlegs(75pt, 95pt, 10pt, 0pt)
\fill ($(D)$) circle (\radius);
\fill ($(B)$) circle (\radius);
\node[anchor = east] at ($0.5*(B) + 0.5*(A) - (1ex,  0)$) {$\tau_2$};
\node[anchor = west] at ($0.5*(D) + 0.5*(C) + (1ex, 0) $) {$\tau_2'$};
\draw [dotted, line width = \lw] ($0.5*(B) + 0.5*(A) + (\radius, 0)$) -- ($0.5*(D) + 0.5*(C) - (\radius, 0)$);
\node[anchor = north] at (\xl + \radius, \yl - 6*\radius) {$i_2$};
\node[anchor = north] at (\xr + \radius, \yl - 6*\radius) {$j_2$};
\end{tikzpicture}
	\caption{Exemplary vertices for the DIM. Diagonal vertices connect two neighboring spins at time $\tau_1=\tau_1'$, while the off-diagonal vertices connect any two spins at time $\tau_2$ and $\tau_2'$. Solid (open) circles symbolize spin up (down).}
\label{fig:operator-vertices}
\end{figure}
The configuration space $C \coloneqq \compbas \times \bigcup_{n=0}^{\infty} \{C_n\}$ is sampled by a modification of the wormhole algorithm following Ref.~\cite{Weber2017, Weber2022b} (see \cite{sm} for details), which is a generalization of the directed loop algorithm \cite{Syljuasen2002} to imaginary-time retarded interactions.
		
In comparison to previous works deploying this wormhole algorithm \cite{Weber2017, Weber2021, Weber2022a, Weber2022b, Weber2024}, we investigate interacting spins coupled to a common bosonic mode. 
As a consequence, a non-local retarded coupling among all pairs of spins is present in \Eq{eq:ret_action} instead of only locally retarded interactions. 
The loop constructed in the directed-loop update can then not only skip over a certain imaginary-time interval \cite{Weber2022b} but, in general, may also jump to another arbitrary site when hitting upon a vertex. 
\begin{figure}[h]
\begin{equation*}
	\Ding \rightarrow \Dbounce + \AFrtraverse  + \sum_{k=1}^{N}\int_0^\beta \dd{\Delta\tau}\left\{\offdiagscatterup + \offdiagscatterdown\right\}
\end{equation*}
\caption{Possible processes when hitting an exemplary diagonal vertex with aligned spins in the directed-loop algorithm for the DIM including an extensive amount of processes and integral in imaginary time. 
The full process visualization leading to the directed-loop equations are shown in \cite{sm}.}
\label{fig:loop-processes}
\end{figure}
The directed-loop equations in this work therefore differ from the commonly encountered ones in the sense that they involve extensive sums over sites and integrals over imaginary time.
The allowed loop processes for an exemplary vertex are illustrated in \Fig{fig:loop-processes}.

In order to identify the different expected phases of the DIM with (anti)ferromagnetic order or superradiance, we measure the (staggered) magnetization $\langle m_{(\rm s)}\rangle$, photon density $\langle\rmi{n}{ph}\rangle$, and energy $\langle\Ham\rangle$ during the simulation. 
While measuring $\langle m_{(\rm s)} \rangle$ is straightforward as $\sz_i$ is diagonal in the computational basis \cite{Sandvik1991, Sandvik2010}, photon observables have to by retrieved indirectly via relations to spin observables \cite{Weber2016, Weber2022b}.
The average photon density $\langle \rmi{n}{ph} \rangle= Z^{-1} \partial{Z}/\partial{\omega}$ can be calculated via
\begin{equation}
\begin{aligned}
	\langle \rmi{n}{ph} \rangle &= \frac{\langle \sum_{k=1}^n \Delta\tau_k \rangle_{\mathrm{MC}}}{\beta} + \frac{e^{-\beta \omega}}{1-e^{-\beta\omega}} \langle 1 + \rmi{n}{od} \rangle_{\mathrm{MC}},
\end{aligned}
\end{equation}
where the first term counts the photons due to the virtual creation of a photon at imaginary time $\tau_k$ and its subsequent annihilation at imaginary time $\tau_k^\prime = \tau_k + \Delta \tau$ by a retarded spin-flip vertex. The second term attributes for thermal photons, which are exponentially suppressed with the photon energy $\omega$ and where $\rmi{n}{od}$ denotes the amount of off-diagonal operators.

The developed wormhole algorithm tailored to treat the non-local light-matter interaction allows us to quantitatively simulate mesoscopic systems with up to 8192 spins. This enables us to determine the quantum phase diagram of the DIM on the chain and the square lattice. However, before discussing the full DIM, we first consider the critical properties and the associated finite-size scaling (FSS) for the limiting case $J=0$ where the model reduces to the well-known Dicke model exhibiting a superradiant phase transition.

{\it Criticality of the Dicke model.} 
Even though the FSS behavior of the Dicke model has been numerically calculated \cite{Lambert2004} in the early 2000s and even determined in a $1/N$ expansion shortly after \cite{Vidal2006}, the origin of the peculiar FSS is not yet well understood and the criticality of the Dicke model has been prone to misconceptions.
In particular, it was commonly found that the shifting and rounding of finite-size observables scale as $N^{-1/\nu'}$ with $\nu'=3/2$ \cite{Lambert2004, Vidal2006, Reslen2005, Liu2009}, which seems incompatible with a $\mathbb{Z}_2$ or mean field symmetry breaking. 
The origin of this FSS and its implication on the criticality of the Dicke model will be resolved in the following.
Further studies in the past also found separate universalities by using the classical instead of the quantum-mechanical susceptibility ($\gamma=1/2$) \cite{Kirton2018} or defining an artificial correlation length $\xi$ ($\nu=1/4$, $z=2$) \cite{Emary2003}.

The criticality of the Dicke model can be understood by linking it to a paradigmatic spin model.
Integrating out the photons in the Dicke model effectively gives a transverse-field Ising model (TFIM) with a normalized all-to-all coupling. 
This can be interpreted as the limit of the long-range TFIM with algebraic decay $\sim \abs{x}^{-(d+\sigma)}$ with decay exponent $d + \sigma = 0$ \cite{Dutta2001, Maghrebi2016, Defenu2017}. 
The long-range TFIM exhibits long-range mean field criticality for $d \geq \min(3, 3\sigma/2)$ with $z\nu = 1/2$, $\beta = 1/2$, and $\gamma = 1$ \cite{Dutta2001, Maghrebi2016, Defenu2017}.
Taking the Dicke model as this limiting case, for which $\duc=0$, the critical exponent $\nu = \infty$ diverges and the space-time anisotropy is maximal with $z=0$ while leaving the gap exponent $z\nu=1/2$ finite.
In the long-range TFIM with all-to-all coupling, the spatial correlations are constant and the correlation length diverges instantaneously when the system orders.

In order to explain the atypical FSS of the Dicke model one has to keep in mind that the model is above the upper critical dimension $\duc$ and the scaling laws are spoiled by dangerous irrelevant variables.
This is taken into account in the Q-FSS approach introduced by R. Kenna \etal for classical systems \cite{Kenna2004} and was extended to quantum systems in Ref.~\cite{Langheld2022}.
In the end, the FSS of observables is modified in the sense that $\nu \rightarrow \nu/\koppa$ and $z \rightarrow z\koppa$ \cite{Langheld2022} with $\koppa = d/\duc$.
This means, the exponent $\nu' = 3/2$ encountered in previous works \cite{Lambert2004, Vidal2006,Liu2009, Reslen2005} is actually $d\nu/\koppa = \nu \duc = 3/2$ as it is the case in the long-range TFIM above the upper critical dimension \cite{Koziol2021, Langheld2022}.
The critical exponents of the Dicke model with long-range mean field criticality are then $\alpha=0$, $\beta=1/2$, $\gamma=1$, $\delta=3$, $\nu=\infty$, and $z\nu=1/2$ \cite{Dutta2001}.
This critical behavior will be present in the full DIM whenever the superradiant ordering is related to a second-order phase transition.

\begin{figure}
	\centering
	\includegraphics{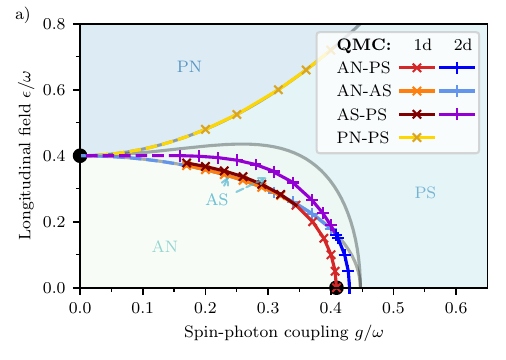}
	\includegraphics{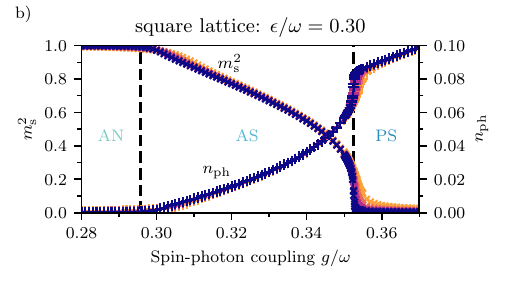}
	\caption{
	Antiferromagnetic DIM for $Jd/\omega=0.2$.  A/P: Antiferro-/Para-magnetic; S/N: Superradiant/Normal. 
	  a) Quantum phase diagram obtained by QMC for the chain (red-toned colors) and square lattice (blue-toned colors). The background shows the phase diagram derived from the variational mean-field ansatz \cite{Zhang2014}. For the chain, the value for $\epsilon = 0$ from self-consistent mean-field is included (see \cite{sm}). The first-order character of the transition found for $\epsilon=0$ \cite{Rohn2020} persists to finite $\epsilon$. b) The order parameters depicted for $\epsilon/\omega = 0.30$ show the existence of the antiferromagnetic superradiant phase for the square lattice with up to $30\cross 30$ (darkblue) spins. }
	\label{fig:phase_diag_aferro}
\end{figure}

{\it Phase diagram of the DIM.} 
\begin{figure}
	\centering
	\includegraphics{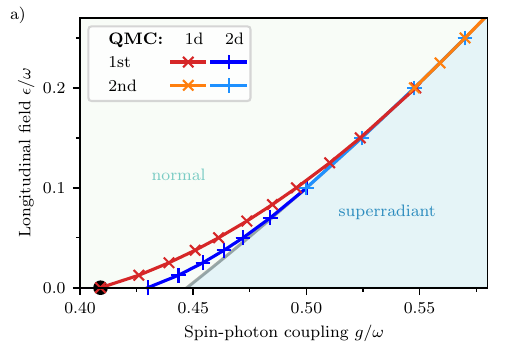}
	\includegraphics{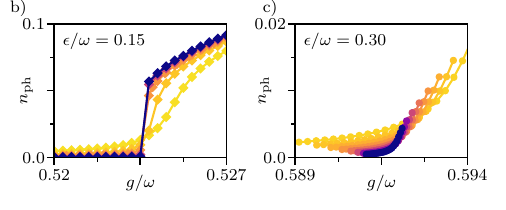}
	\caption{Ferromagnetic DIM for $Jd/\omega=-0.2$. a) Quantum phase diagram obtained by QMC for the chain (red-toned colors) and square lattice (blue-toned colors). The background corresponds to the results from the variational mean-field ansatz valid in infinite dimensions.  For the chain, the exact value for $\epsilon = 0$ from self-consistent mean-field approach is included (see \cite{sm}). For small $\epsilon \ll \abs{J}$, the transition is of first order, for large $\epsilon \gg \abs{J}$, the transition is of second order and shows Dicke criticality. Both regimes are separated by a multicritical point.
	b) Order parameter $n_{\rm ph}$ for up to 4096 (darkblue) spins for $\epsilon/\omega=0.15$ where the transition is of first order. c) Order parameter $n_{\rm ph}$ for up to 8192 (darkblue) spins for $\epsilon/\omega=0.3$ where the transition is of second order.}
	\label{fig:phase_diag_ferro}
\end{figure}
The obtained quantum phase diagrams for ferro- and antiferro-magnetic Ising interactions are displayed in \Fig{fig:phase_diag_ferro} and  \Fig{fig:phase_diag_aferro} respectively (see \cite{sm} for the effect of an \mbox{$A^2$-term}).
As the $\mathbb{Z}_2$ spin-flip symmetry is broken for any longitudinal field $\epsilon$, there is only one transition from the normal to the superradiant phase in the ferromagnetic case.
Applying the same variational mean field ansatz as in \cite{Zhang2014}, one finds a second-order transition at $\epsilon = 2(g^2/\omega - \abs{J}d)$, which is the same expression as for the antiferromagnetic case (see background in \Fig{fig:phase_diag_ferro}).
In contrast, our unbiased QMC results for the chain and the square lattice show that this transition is first order for small longitudinal fields $\epsilon \ll \abs{J}$ in agreement with self-consistent mean field \cite{Puel2024}, while for large longitudinal fields $\epsilon \gg \abs{J}$, the transition becomes second order. In the latter regime the QMC results match perfectly with the ones from the variational mean field ansatz which is consistent with recent analytical findings \cite{Schellenberger2024}. 
The second-order transition exhibits the same criticality (long-range mean field) as the pure Dicke model with $J=0$ as discussed above. 
The first- and second-order regimes connect at a multicritical point at approximately $\epsilon \approx \abs{J}$ which shifts to smaller values of $\epsilon$ for constant $\abs{J}d$ and increasing spatial dimension.

For antiferromagnetic Ising interactions, the translational symmetry of the lattice and the $\mathbb{Z}_2$ symmetry of the Dicke model can be broken by antiferromagnetic order and superradiance respectively. 
This leads to four quantum phases obeying distinct symmetries.
In \Fig{fig:phase_diag_aferro}, we display the quantitative ground-state phase diagram for the chain and the square lattice from QMC together with the qualitative predictions from the variational mean field ansatz from Ref.~\cite{Zhang2014}. 
The first-order transition point at $\epsilon = 0$ from antiferromagnetic normal to paramagnetic superradiant phase, with one type of order emerging while the other vanishes, continues as phase transition line at finite longitudinal fields until it crosses the mean field transition line between the antiferromagnetic normal and antiferromagnetic superradiant phase.
Here it splits into two separate transition lines.
By this, the superradiance already occurs for smaller spin-photon couplings $g$ than the breakdown of antiferromagnetic order which leads to the formation of the intermediate antiferromagnetic superradiant phase with both types of order simultaneously.
As expected, the second-order transition lines from antiferromagnetic (paramagnetic) normal to antiferromagnetic (paramagnetic) superradiant phase have the same universality class (long-range mean field) as the pure Dicke model with $J=0$. They are again in perfect agreement with the predictions from the variational mean field ansatz \cite{Zhang2014}, already for 1d. We therefore focused the computational ressources for 2d on the more interesting regime of the intermediate phase and its phase boundaries. 

For the square lattice, the non-trivial transition line between a magnetically-ordered and a paramagnetic superradiant phase, which emerges from the first-order transition line for small $\epsilon$, remains of first order for $g > 0.30(2)$ but changes to a second-order quantum phase transition with 3D Ising criticality for $g < 0.30(2)$.
The criticality of the second-order phase transition is in agreement with \cite{Kaneko2021} considering the antiferromagnetic TFIM in a longitudinal field on the square lattice, which found 3D Ising universality for any finite transverse field.
For the linear chain, similar studies exist observing 2D Ising criticality \cite{Bonfim2019, Ovchinnikov2003}.
However, our QMC simulations for the linear chain showed no indication that the transition between the antiferro- and paramagnetic superradiant phase becomes continuous at any point in the phase diagram.

{\it Conclusions.} 
In this work we designed a wormhole algorithm which is tailored for the paradigmatic DIM combining light-matter and matter-matter interactions. 
This allows us to quantitatively enter the domain of mesoscopic light-matter systems relevant for a variety of experimental platforms at the interface of quantum optics and condensed matter physics.
Here we determined the rich quantum phase diagram of the DIM containing first- and second-order phase transitions as well as multicritical points in a quantitative fashion for ferromagnetic and antiferromagnetic Ising interactions in one and two dimensions. 
All second-order superradiant phase transitions are in quantitative agreement with variational mean field theory \cite{Zhang2014} valid in infinite dimensions which is in line with recent analytical findings \cite{Schellenberger2024}. 
These second-order phase transitions turned out to lie in the same universality class as the pure Dicke model. 
Alongside our analysis of these phase transitions, we addressed misconceptions on the criticality of the Dicke model in previous works \cite{Kirton2018, Emary2003}, and put its peculiar FSS \cite{Vidal2006} into line with the expected criticality of the model by means of Q-FSS \cite{Kenna2004, Langheld2022}. 
Most importantly, we confirm the presence of an intermediate phase with simultaneous antiferromagnetic and superradiant order for the DIM on the chain and the square lattice.
The extension of this phase, however, is much smaller than predicted by the mean field limit \cite{Zhang2014}. 
The antiferromagnetic superradiant phase can be viewed as the light-matter analogue of a lattice supersolid known from interacting quantum many-body systems. 

Our algorithm is expected to give valuable insights well beyond the studied DIM on the bipartite chain and square lattice.
It can easily be generalized to a continuum of modes \cite{Weber2022b} or adapted to locally different light-matter couplings in the DIM.
This will allow to investigate the effects of bosonic modes with finite momenta beyond the long wavelength approximation or the effects of disorder.   
Further, our algorithm is sign-problem free for DIMs with geometrically frustrated Ising interactions. 
This paves the way to explore frustrated light-matter systems which we expect to stabilize novel light-matter phenomena in the quantum domain \cite{Schuler2020}.

\begin{acknowledgments}
We thank J. A. Koziol, C. Kr\"amer and A. Schellenberger for valuable input and fruitful discussions.
	The authors gratefully acknowledge the support by the Deutsche Forschungsgemeinschaft (DFG, German Research Foundation) - Project-ID 429529648 - TRR 306 \mbox{QuCoLiMa} ("Quantum Cooperativity of Light and Matter") and the Munich Quantum Valley, which is supported by the Bavarian state government with funds from the Hightech Agenda Bayern Plus. We gratefully acknowledge the scientific support and HPC resources provided by the Erlangen National High Performance Computing Center (NHR@FAU) of the Friedrich-Alexander-Universit\"at Erlangen-N\"urnberg (FAU). The hardware is funded by the German Research Foundation (DFG).
\end{acknowledgments}

\subsection*{Data availability}
The QMC data as well as the numerical results extracted from it for this work (points of phase transition, critical exponents $\nu$ and $\beta$ where applicable) are openly available \cite{RawData}.

%insert bibliography here
%

\renewcommand{\theequation}{S\arabic{equation}}
\renewcommand{\thefigure}{S\arabic{figure}}
\pagebreak
\widetext
\newcommand{\titlepaper}{}
\begin{center}
	\textbf{\large Supplemental Material for ``Quantum phase diagrams of Dicke-Ising models by a wormhole algorithm''}
	\\\vspace{0.5em}
	Anja Langheld$^{\rm 1}$, Max H\"ormann$^{\rm 1}$ and Kai P. Schmidt$^{\rm 1}$ \\
	\textit{$^{\text 1}$Friedrich-Alexander-Universit\"at
		Erlangen-N\"urnberg (FAU), \\Department Physik,
		Staudtstra{\ss}e 7, D-91058 Erlangen, Germany}
\end{center}

\setcounter{equation}{0}
\setcounter{figure}{0}
\setcounter{table}{0}
\setcounter{page}{1}
\makeatletter
\renewcommand{\theequation}{S\arabic{equation}}
\renewcommand{\thefigure}{S\arabic{figure}}
\renewcommand{\bibnumfmt}[1]{[S#1]}
\renewcommand{\citenumfont}[1]{S#1}
\setcounter{secnumdepth}{3}
\setcounter{section}{0}

In this supplemental material we describe in depth the methods we employed to create the data presented in the main text. In particular, in \Sec{app:algo} we explain the wormhole Quantum Monte Carlo method \cite{Weber2022bs} for the case of the Dicke-Ising model. 
We provide a solution to the directed-loop equations, the data structures and data analysis tools that we employed and details about how the simulations were conducted. 
These detailed explanations might be helpful for readers who would like to implement this algorithm.
In order to decrease the barrier to sight the raw data, we discuss certain physical results of the main paper like the order of certain phase transitions, multicritical points or criticality of second-order phase transition on the basis of representative raw datasets. All raw datasets are made available in \cite{RawDatas} for individual inspection.
Furthermore, in \Sec{app:sc-mf} we outline the self-consistent mean field calculation to derive the exact value of the quantum phase transition for the Dicke-Ising model on the linear chain with $\epsilon = 0$.
This value was included in the phase diagrams of the main part and within this supplemental material we use it to demonstrate the excellent accuracy of the wormhole algorithm employed. 
Moreover, we present the effect of an additional diamagnetic $A^2$-term in the Hamiltonian on the quantum phase diagram of the Dicke-Ising model in \Sec{app:A-squared}. The incorporation of this term leads to a boundary line in parameter space separating the accessible region of the phase diagrams from the inaccessible one.

\section{Wormhole algorithm for simulating Dicke-Ising models}

\label{app:algo}

Considering a spin-boson model which is quadratic in the bosonic operators, one can integrate out the bosons giving an effective spin model with the partition function \cite{Weber2022bs}
\begin{equation}
	\label{eq:ret-part-func}
	Z = \rmi{Z}{b} \rmi{\Tr}{s}\left[ \mathcal{T}_{\tau} e^{-\rmi{S}{ret}} \right]
\end{equation}
 with $\rmi{Z}{b}$ the partition function of free bosons, $\mathcal{T}_{\tau}$ the time-ordering operator and $\rmi{\Tr}{s}\left[\dots\right]$ the spin trace with a retarded effective spin-spin action $\rmi{S}{ret}$. The retarded spin-spin action for the Dicke-Ising model is given by
\begin{equation}
	\begin{aligned}
		\rmi{S}{ret} &= \int_0^\beta \dd{\tau} \left[ \epsilon \sum_{i=1}^N \sz_i(\tau) + \frac{J}{2}\sum_{i = 1}^N \sum_{j \in \mathrm{nn}(i)} \sz_i(\tau)\sz_j(\tau)\right] - 
		\frac{g^2}{\omega N} \int_0^\beta \dd{\tau}\int_0^\beta \dd{\tau'} \sum_{i,j=1}^N \sx_i(\tau) P(\omega, \tau - \tau') \sx_j (\tau')
	\end{aligned}
\end{equation}
where $P(\omega, \Delta\tau) = \omega \frac{e^{-\Delta\tau \omega}}{1- e^{-\beta\omega}}$ is the (normalized) free boson propagator. 
The sum over nearest-neighbor spin pairs $\langle i,j \rangle$ is rewritten as a sum over all spins $i \in \{1,\dots, N\}$ and sum over all nearest neighbors $j \in \mathrm{nn}(i)$ of spin $i$ for algorithmic purposes.
This leads to a double counting of all Ising bonds which is compensated for by introducing a factor of $1/2$ to the bond strength $J$.
As the retarded interaction is symmetric in $\tau$ and $\tau'$, one can also use the symmetrized free boson propagator \cite{Weber2022bs}
\begin{align}
	\rmi{P}{sym} (\omega, \Delta\tau) &= \frac{P(\omega, \Delta \tau) + P(\omega, \beta - \Delta\tau)}{2}
\end{align}
to treat $\tau$ and $\tau'$ on equal footing.

In order to sample the partition function with the approach of stochastic series expansion (SSE), one first needs to choose a computational basis $\compbas$ in which the spin trace $\Tr_s\left[\dots\right]=\sum_{\compbas} \bra{\alpha}\dots\ket{\alpha}$ is evaluated.
For the Dicke-Ising model, we have chosen the $\sz$-basis such that the Ising interaction and longitudinal field are diagonal in the computational basis.
Further, one needs to decompose the action $\rmi{S}{ret} =  -\SumInt_{\nu} \Ham_\nu\dd{\tau}\dd{\tau'}$ into a sum of operators $\Ham_\nu$ specified by a multi-index $\nu$. The decomposition needs to be chosen such that all $\Ham_\nu$ are non-branching in the computational basis $\compbas$ in which the spin trace $\Tr_s\left[\dots\right]$ is evaluated \cite{Sandvik1991s, Sandvik2010s}. This means that they do not create superpositions of basis states when acting upon a basis state.
The chosen decomposition for the Dicke-Ising model is
\begin{align}
	\rmi{S}{ret} &= - \int_0^\beta \int_0^\beta \dd{\tau}\dd{\tau'} \left[\sum_{i=1}^N \sum_{j\in\mathrm{nn}(i)}\Ham^{\mathrm{d}}_{ij}(\tau, \tau') +  \sum_{i,j=1}^N\Ham^{\mathrm{od}}_{ij}(\tau, \tau')\right]
\end{align}
with $\Ham^{\mathrm{d}}$ being operators diagonal in the computational basis and $\Ham^{\mathrm{od}}$ being the off-diagonal retarded operators. 
The multi-index $\nu$ is therefore given by $\nu=\{\mathrm{type}, i, j, \tau, \tau'\}$ with the $\mathrm{type}$ being diagonal (d) or off-diagonal (od), $i$ and $j$ being the pairs of sites on which the operator acts on and $\tau, \tau'\in[0,\beta)$ the imaginary times at which the operator acts. 
Explicitly, the operators $\Ham^{\mathrm{d/od}}_{ij}(\tau,\tau')$ are given by
\begin{align}
	\Ham^{\mathrm{d}}_{ij}(\tau, \tau') &= \frac{\delta_{\tau, \tau'}}{2} 
		\left[\abs{J}-J\sz_i(\tau) \sz_j (\tau) + \frac{\epsilon}{z}(2- \sz_i(\tau) - \sz_j(\tau))\right]\\
	\Ham^{\mathrm{od}}_{ij}(\tau, \tau') &= \frac{g^2}{\omega N} \sx_i(\tau) \rmi{P}{sym}(\omega, \tau - \tau') \sx_j (\tau') \label{eq:off-diag-op}
\end{align}
acting on pairs of spins $i,j$ with $z$ being the number of neighbors per site.
The constant shift of $\abs{J}/2 + \epsilon/z$ added to the diagonal operators ensures that all the matrix elements in the computational $\sz$-basis are non-negative to avoid the sign problem. 
This leads to an overall constant added to $\rmi{S}{ret}$ which only amounts to a constant shift in the energy spectrum but does not alter the (eigen)states or their properties aside from their energy.
One just needs to make sure to subtract this constant when calculating observables like the mean energy $\langle \Ham \rangle$ of the system.

The partition function \Eq{eq:ret-part-func} is then rewritten as in the SSE approach \cite{Sandvik1991s, Sandvik2010s} 
\begin{align}
	\frac{Z}{Z_b} &= \sum_{\compbas} \sum_{n=0}^{\infty} \frac{1}{n!} \bra{\alpha} \mathcal{T}_\tau  \rmi{S}{ret}^n\ket{\alpha}\\
	&= \sum_{\compbas} \sum_{n=0}^{\infty} \sum_{\{C_n\}}\frac{1}{n!} \bra{\alpha} \mathcal{T}_\tau  \prod_{k=1}^n \rmi{\Ham}{\nu_k}\dd{\tau_k}\dd{\tau'_k}\ket{\alpha} \label{eq:part-func-expansion}%\\
\end{align}
where $\mathcal{T}_\tau$ denotes the time-ordering operator acting on all $\tau_k,\tau_k'$ together, $C_n=\{\Ham_{\nu_1}, \dots, \Ham_{\nu_n}\}$ denotes the ordered vertex list with vertex variables $\nu_k = \{\mathrm{type}_k, i_k, j_k, \tau_k, \tau_k'\}$ for every operator $\Ham_{\nu_k}$. 
In order to further rewrite the partition function into a sum of weights, one needs to get rid of the operator structure, including the time-ordering operator $\mathcal{T}_\tau$. 
For this, one needs to split the operators $\Ham_{\nu_k} = A_{\nu_k}(\tau_k') B_{\nu_k}(\tau_k)$ into operators $A_{\nu_k}(\tau_k')$ and $B_{\nu_k}(\tau_k)$ that act at imaginary times $\tau_k'$ and $\tau_k$ respectively. 
If an operator $\Ham_{\nu_k}$ has a global amplitude, \eg, ${\Ham_{ij}^{\mathrm{od}}(\tau, \tau') \sim g^2/(\omega N) \rmi{P}{sym}(\omega, \tau_k'-\tau_k)}$ for off-diagonal operators (see \Eq{eq:off-diag-op}), it does not matter how it is absorbed into $A$ or $B$ as both operators will only contribute together to the total weight of the configuration. 
For now, it is only important to split the action of the operators at times $\tau_k$ and $\tau'_k$ of every operator $\Ham_{\nu_k}$ so one can properly define the (normalized) propagated state
\begin{align}
	\ket{\alpha(\tau)} \sim \mathcal{T}_{\tau} \prod_{\tau_k' < \tau} A_{\nu_k}(\tau_k')\prod_{\tau_k < \tau}B_{\nu_k}(\tau_k) \ket{\alpha}\, .
\end{align}
In words, the state $\ket{\alpha(\tau)}$ is simply the normalized state after applying all operators in the time-ordered operator sequence $\mathcal{T}_\tau  \prod_{k=1}^n \rmi{\Ham}{\nu_k}$ until time $\tau$ onto the reference basis state $\ket{\alpha}$ at $\tau=0$.
We stress that the state $\ket{\alpha(\tau)}$ does not refer to a state after proper evolution in imaginary time as generated by $\exp(-\tau \Ham)$, but it is only a vertex-list dependent portion of a proper imaginary-time evolution.

With this propagated state, \Eq{eq:part-func-expansion} can be rewritten as matrix elements of these propagated states
\begin{align}
	\frac{Z}{Z_b} &= \sum_{\compbas} \sum_{n=0}^{\infty} \sum_{\{C_n\}}\frac{1}{n!} \prod_{k=1}^n \lim_{\epsilon \rightarrow 0^+}\bra{\alpha (\tau'_k + \epsilon)}  A'_{\nu_k}\dd{\tau_k'}\ket{\alpha(\tau_k')} \bra{\alpha (\tau_k + \epsilon)} B_{\nu_k}\dd{\tau_k}\ket{\alpha(\tau_k)}\\
	&= \sum_{\compbas} \sum_{n=0}^{\infty} \sum_{\{C_n\}} w(\alpha, C_n)\, .
\end{align}
The sum can then be sampled by a Markovian random walk within the configuration space $\mathcal{C}\coloneqq \compbas \times \bigcup_{n=0}^{\infty} \{C_n\}$ with stationary distribution proportional to the relative weights
\begin{align}
	w(\alpha, C_n) &= \frac{1}{n!} \prod_{k=1}^n \lim_{\epsilon \rightarrow 0^+}\bra{\alpha (\tau'_k + \epsilon)}  A'_{\nu_k}\dd{\tau_k'}\ket{\alpha(\tau_k')} \bra{\alpha (\tau_k + \epsilon)} B_{\nu_k}\dd{\tau_k}\ket{\alpha(\tau_k)}\, .
\end{align}
Due to the non-branching property, all of the propagated states $\ket{\alpha(\tau_k)}$ and $\ket{\alpha(\tau^\prime_k)}$, on which the operators act upon, are ensured to be basis states of the computational basis $\compbas$. 
Hence, the matrix elements appearing in the weights can only take on a few values depending on the local spin configuration at the sites $i_k, j_k$ and imaginary times $\tau_k,\tau_k'$.
All non-vanishing matrix elements of the operators in the computational basis can be represented by a vertex that connects the spin states of a pair of spins before and after applying the respective operator.
A local vertex is therefore defined by four spin states, namely the state of two spins before and after applying the operator.
The weight of a vertex is given by the value of its defining matrix element.
Several vertices can have the same weight and one can therefore group them into vertex categories.
The different possible vertex categories with non-vanishing matrix elements are illustrated in \Fig{fig:operators} together with their respective weights.

\begin{figure}
\begin{center}
\begin{tikzpicture}
	\offdiagophalf(225pt, 245pt, 10pt, 0pt)
	\diagopd(0pt, 0pt)
	\node[anchor = center] at (\xr, \y - 12*\radius) {$\rmi{w}{od}(\Delta \tau)$};
	\node[anchor = center] at (\xr, \y - 24*\radius) {$\frac{g^2}{\omega N}\rmi{P}{sym}(\omega, \Delta \tau)$};
	\node[anchor = west] at (\x - 70pt + 0.5*\dx + \radius, \y + 20*\radius) {\textbf{diagonal:}};
	\node[anchor = west] at (\xr - 70pt + 0.5*\dx + \radius, \y + 20*\radius) {\textbf{off-diagonal:}};
	\node[anchor = center] at (\x - 40pt + 0.5*\dx + \radius, \y - 20*\radius) {ferro};
	\node[anchor = center] at (\x - 40pt + 0.5*\dx + \radius, \y - 28*\radius) {antiferro};
	\node[anchor = center] at (\x + 0.5*\dx + \radius, \y - 20*\radius) {$\abs{J}$};
	\node[anchor = center] at (\x + 0.5*\dx + \radius, \y - 28*\radius) {$0$};
	\node[anchor = center] at (\x + 0.5*\dx + \radius, \y - 12*\radius) {$\rmi{w}{up}$};
	\def\xAFl{42pt}
	\def\xAFr{84pt}
	\def\xAFl{48pt}
	\def\xAFr{78pt}
	\diagopafr(\xAFl, 0pt)
	\diagopaf(\xAFr, 0pt)
	\node[anchor = center] at (0.5*\xAFl + 0.5*\xAFr + 0.5*\dx + \radius, \y - 12*\radius) {$\rmi{w}{AF}$};
	\node[anchor = center] at (0.5*\xAFl + 0.5*\xAFr + 0.5*\dx + \radius, \y - 20*\radius) {$\frac{\epsilon}{z}$};
	\node[anchor = center] at (0.5*\xAFl + 0.5*\xAFr + 0.5*\dx + \radius, \y - 28*\radius) {$\abs{J} + \frac{\epsilon}{z}$};
	\diagopu(126pt, 0pt)
	\node[anchor = center] at (\x + 0.5*\dx + \radius, \y - 20*\radius) {$\abs{J} + 2\frac{\epsilon}{z}$};
	\node[anchor = center] at (\x + 0.5*\dx + \radius, \y - 28*\radius) {$2\frac{\epsilon}{z}$};
	\node[anchor = center] at (\x + 0.5*\dx + \radius, \y - 12*\radius) {$\rmi{w}{down}$};
\end{tikzpicture}
\end{center}
	\caption{Visualization of the different diagonal and off-diagonal vertices and their respective weights $\rmi{w}{up,AF,down}$ and $\rmi{w}{od}$ determined by the matrix element of the spin state ($\uparrow = \bullet$, $\downarrow = \circ$) they act upon. In general, there are four non-vanishing matrix elements per off-diagonal operator $H_{ik}^{\mathrm{od}}(\tau, \tau')$ and up to four non-vanishing matrix elements per diagonal operator $H_{ij}^{\mathrm{d}}(\tau, \tau)$.}
	\label{fig:operators}
\end{figure}

As commonly done in the SSE scheme \cite{Sandvik1991s, Sandvik2010s}, one step in the Markov chain is subdivided into two distinct update procedures; one diagonal update in which diagonal operators are inserted and removed from the ordered vertex list $C_n$ and one off-diagonal update in which diagonal operators are transformed into off-diagonal operators and vice versa.

The diagonal update consists of a certain amount $\rmi{m}{diag}$ of proposed insertions and removals of diagonal operators in the ordered vertex list $C_n$. 
An operator is attempted to be inserted or removed with probability $1/2$ respectively.
In total, there are $N\cdot z$ diagonal operators that can be positioned at an arbitrary imaginary time $\tau\in[0,\beta)$ and at $n+1$ different positions in the ordered vertex list. 
Drawing a diagonal operator $\Ham^{\mathrm{d}}_{ij}(\tau, \tau' = \tau)$ uniformly in these vertex variables and an insertion position $k$ uniformly in the ordered vertex list gives the proposal probability
\begin{align}
	\rmi{q}{in}(\Ham^{\mathrm{d}}_{ij}(\tau, \tau), k) = \frac{1}{Nz} \frac{\dd{\tau}}{\beta} \frac{1}{n+1}
\end{align}
to propose this particular operator at a certain position in $C_n$.
For the reverse process, there are $\rmi{n}{diag}$ possibilities to choose a diagonal operator $\Ham^{\mathrm{d}}_{ij}(\tau, \tau)$ from the ordered vertex lists $C_n$. Drawing a diagonal operator at index $k$ uniformly from all diagonal operators in $C_n$ is therefore done with the proposal probability
\begin{align}
	\rmi{q}{rem}(\Ham^{\mathrm{d}}_{ij}(\tau, \tau), k) = \frac{1}{\rmi{n}{diag}}\, .
\end{align}
After proposing a certain diagonal operator $\Ham_\nu$ for insertion or removal at vertex-list index $k$, the move is accepted with probability $\rmi{p}{in}(\Ham_{\nu}, k)$ for inserting and $\rmi{p}{rem}(\Ham_{\nu},k)$ for removing an operator chosen according to the Metropolis-Hastings algorithm as
\begin{align}
	\rmi{p}{in} (\Ham_\nu, k) &= \min\left(1, \frac{ \bra{\alpha(\tau)} \Ham_\nu \ket{\alpha(\tau)}\beta N z}{\rmi{n}{diag} + 1}\right)\\
	\rmi{p}{rem} (\Ham_\nu, k) &= \min\left(1, \frac{\rmi{n}{diag}}{ \bra{\alpha(\tau)} \Ham_\nu \ket{\alpha(\tau)}\beta N z}\right)\, .
\end{align}

After $\rmi{m}{diag}$ insertions/removals have been proposed, the off-diagonal update is conducted, in which the operator types as well as the basis state $\ket{\alpha}$ can change.
The off-diagonal update procedure consists of $\rmi{m}{worm}$ wormhole updates, which are updates that extend the directed-loop algorithm \cite{Syljuasen2002s} to systems with retarded interactions as introduced in Ref.~\cite{Weber2017s}. 
The key difference in contrast to the regular directed-loop update is that the loop in the wormhole update can have a discontinuity in imaginary time and continue at a different site and imaginary time due to the non-locality of the retarded interaction vertices.
However, the construction of the directed-loop equations is analogous \cite{Weber2017s}. 

\begin{figure}
\begin{equation*}
	\left.
	\begin{aligned}
		\Ding &\rightarrow \Dbounce + \AFrtraverse  + \sum_{k=1}^{N}\int_0^\beta \dd{\Delta\tau}\left\{\offdiagscatterup + \offdiagscatterdown\right\}\\
		\AFingd &\rightarrow \AFbounced + \Dtraverse + \sum_{k=1}^{N}\int_0^\beta \dd{\Delta\tau}\left\{\offdiagscatterupfromup + \offdiagscatterdownfromup\right\}\\
		\AFing &\rightarrow \AFbounce + \Utraverse  + \sum_{k=1}^{N}\int_0^\beta \dd{\Delta\tau}\left\{\offdiagscatterup + \offdiagscatterdown\right\}\\
		\Uingd & \rightarrow \Ubounce + \AFtraverse + \sum_{k=1}^{N}\int_0^\beta \dd{\Delta\tau}\left\{\offdiagscatterupfromup + \offdiagscatterdownfromup\right\}
	\end{aligned}
	\quad
	\right\vert
	\quad
	\begin{aligned}
		\offdiaging & \rightarrow \offdiagbounce + \offdiagtraverse + \sum_{j\in \mathrm{nn}(i)}
		\begin{cases}
			\Dscatter + \AFscatterx \text{\ \ if $j$ is } \fullcircle  \\
			\AFscattero + \Uscatter \text{\ \ if $j$ is } \emptycircle 
		\end{cases}
		\\
		\offdiagingup & \rightarrow \offdiagbounceup + \offdiagtraverseup + \sum_{j=\in \mathrm{nn}(i)}
		\begin{cases}
			\Dscatter + \AFscatterx \text{\ \ if $j$ is } \fullcircle  \\
			\AFscattero + \Uscatter \text{\ \ if $j$ is } \emptycircle 
		\end{cases}
	\end{aligned}
\end{equation*}
	\caption{Visualization of the allowed processes for the loop construction when hitting the different vertex types during the loop construction. In the directed-loop equations the weight of the vertex on the left-hand side of $\rightarrow$ has to equal the flow into the vertices on the right-hand side of $\rightarrow$. }
\label{fig:worm-scattering}
\end{figure}

For the Dicke-Ising model, the considered vertex-conversion processes for the loop construction, which compose the directed-loop equations, are shown in \Fig{fig:worm-scattering} for the different vertices.
In contrast to the previous applications of the wormhole algorithm, we consider a spin-boson system with a common bosonic mode alongside a lattice-based spin-spin interaction.
These two interaction types give rise to very dissimilar spin-spin interactions and consequently very distinct operator vertices. While the diagonal operators only connect neighboring spins, off-diagonal operators connect all pairs of spins and additionally have a weight that decreases with the amount of spins with $ 1/N$. This makes it more complicated to construct an algorithm which efficiently converts one type of operators into the other.

For the algorithm used in the main part, we therefore allow diagonal bonds at neighboring sites $i,j$ at a certain imaginary time $\tau$ to transform into an off-diagonal operator at sites $i, k$ or $k, j$ with $k\in{1,\dots, N}$ and at imaginary time $\tau$ and $\tau'$ and vice versa.
When transforming a diagonal operator into an off-diagonal operator, the new site $k$ and new imaginary time $\tau'$ are drawn during the wormhole algorithm and are not predetermined by the diagonal operator that is being transformed, also in contrast to earlier implementations \cite{Weber2017s, Weber2022bs}. The same holds for the new neighboring site $j$ (w.\,l.\,o.\,g.) when transforming an off-diagonal operator at sites $i, k$ into a diagonal bond at $i,j$.

This gives rise to directed-loop equations that include sums over all spins $1,\dots,N$ and integrals over imaginary time as visible in \Fig{fig:worm-scattering}. This is different from common directed-loop algorithms in which only a few conversion processes are allowed (usually around four: bounce, straight, switching sites and U-turn). 
The vast number of allowed processes in our case not only potentially helps the update to better shuffle the different kind of operators, but allowing diagonal operators to be transformed into an extensive amount of off-diagonal operators will compensate for the $1/N$-factor in the off-diagonal vertex weights in the directed-loop equations.
By this, the probability to convert a diagonal operator into an off-diagonal operator and vice versa is constant in $N$.

Based on the conversion processes shown in \Fig{fig:worm-scattering}, one can set up local equations for the loop construction. As it is commonly the case, we choose them to satisfy the detailed balance condition,
\begin{equation}
	\left.
	\begin{aligned}
	\rmi{w}{up} &= b_{\bullet\bullet} + a + \sum_{k=1}^{N}\int_0^\beta \dd{\Delta\tau}e(\Delta\tau) + e'(\Delta\tau)\\
	\rmi{w}{AF} &= b_{\bullet\circ} + a + \sum_{k=1}^{N}\int_0^\beta \dd{\Delta\tau}f(\Delta\tau) + f'(\Delta\tau)\\
	\rmi{w}{AF} &= b_{\circ\bullet} + c + \sum_{k=1}^{N}\int_0^\beta \dd{\Delta\tau}g(\Delta\tau) + g'(\Delta\tau)\\
	\rmi{w}{down} &= b_{\circ\circ} + c + \sum_{k=1}^{N}\int_0^\beta \dd{\Delta\tau}h(\Delta\tau) + h'(\Delta\tau)\\
	\end{aligned}
	\qquad
	\right\vert
	\qquad
	\begin{aligned}
	\rmi{w}{od} (\Delta\tau) &= \rmi{b}{od} + d + \sum_{j \in \mathrm{nn}(i)}
	\begin{cases}
		e + f \text{\ \ if $j$ is } \fullcircle  \\
		g + h \text{\ \ if $j$ is } \emptycircle 
	\end{cases}\\
	\rmi{w}{od} (\Delta\tau) &= \rmi{b}{od}' + d + \sum_{j \in \mathrm{nn}(i)}
	\begin{cases}
		e' + f' \text{\ \ if $j$ is } \fullcircle  \\
		g' + h' \text{\ \ if $j$ is } \emptycircle  \, ,
	\end{cases}
	\end{aligned}
	\label{eq:wormhole-detailed-balance}
\end{equation}
where $\rmi{w}{up,AF,down}$ and $\rmi{w}{od}$ are the weights of the vertices as defined in \Fig{fig:operators}. These weights are a sum of different process weights. For instance, $b_{\mathrm{xx}}$ are the bounce process weights, \eg, $\rmi{b}{\bullet \bullet} = \rmi{w}{up} \rmi{p}{up,b}$ with $\rmi{p}{up,b}$ the probability for a loop to bounce when entering the vertex $(\Diagopd)$ with all spins pointing up. 
The weights $a,c,d$ are the weights of the processes in which the loop traverses the vertex straight and the weights $e,f,g,h$ and their primed counterparts correspond to the conversion processes of diagonal to off-diagonal operators and vice versa.

We simplify these equations by several plausible choices. All the loop processes were already taken to be independent of the involved spin sites $i,j$ and $k$ based on the translational invariance of the underlying model. One can therefore substitute $\sum_{k=1}^N \rightarrow N$ and $\sum_{j\in \mathrm{nn}(i)} \rightarrow z$.
Further, in the process of replacing a diagonal operator with an off-diagonal operator, the probability for the loop to continue in positive or negative direction in imaginary time after exiting the new off-diagonal operator should be equal as all off-diagonal vertices have the same weight independent of the spin state they act on.
Therefore we set $x=x'$ with $x=e,f,g,h$. This leads to a factor of 2 when rewriting $x+x'=2x$ and makes the last two directed-loop equations for the off-diagonal vertex equivalent.
Moreover, we take $e+f=g+h$ to avoid the need to distinguish between the states of the neighboring sites $j\in \mathrm{nn}(i)$ being $\bullet$ or $\circ$ when transforming an off-diagonal vertex at sites $i,k$ to a diagonal vertex at sites $i,j$.
Otherwise one would need to check the spin states of all sites $j$ that are next to site $i$ when an off-diagonal vertex connecting sites $i$ and $k$ is entered by the loop head at site $k$.
Furthermore, it is a reasonable choice to take $x(\Delta\tau)\sim \rmi{P}{sym}(\omega, \Delta\tau)$ such that the proposal to convert a diagonal operator into an off-diagonal operator already respects the relative weight $\rmi{w}{od}(\Delta\tau)\sim \rmi{P}{sym}(\omega, \Delta\tau)$ of the off-diagonal operator. 
This makes the bounce, traverse and scattering probabilities when entering an off-diagonal vertex independent of $\Delta\tau$.
Additionally defining $X = N \int \dd{\Delta \tau} x(\Delta\tau)$ for $X=E,F,G,H$ ($x=e,f,g,h$) respectively, $\rmi{W}{od} = N\int_0^\beta \rmi{w}{od}\dd{\Delta\tau} = g^2/\omega$ and analogously $\rmi{B}{od}$ and $D$ in terms of $\bod$ and $d$, one arrives at these simplified equations
\begin{equation}
	\begin{split}
	\label{eq:worm-equation}
	\rmi{w}{up} &= b_{\bullet\bullet} + a + 2E\\
	\rmi{w}{AF} &= b_{\circ\bullet} + a + 2F\\
	\rmi{w}{AF}&= b_{\bullet\circ} + c + 2G\\
	\rmi{w}{down} &= b_{\circ\circ} + c + 2H\\
	\rmi{W}{od} &= \rmi{B}{od} + D + z\cdot (E+F)\\
	\text{ with } &\quad E + F = G + H\, .
	\end{split}
\end{equation}

\subsection{Solution and implementation of directed-loop equations}
\begin{figure}
	\includegraphics{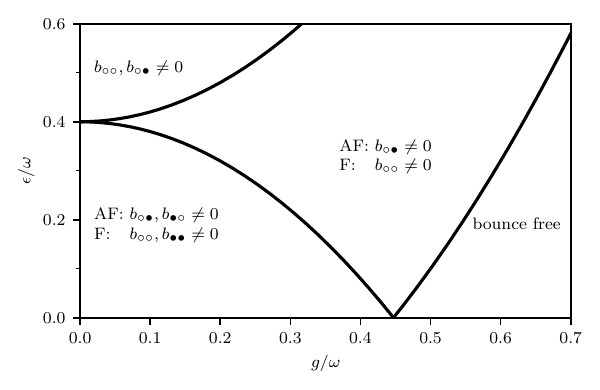}
	\caption{Algorithmic phase diagram for $\abs{J}z/\omega=0.2$ as a function of $g/\omega$ and $\epsilon/\omega$ analogous to the physical phase diagram in the main text. The phase boundaries are the same for ferromagnetic (F) and antiferromagnetic (AF) Ising couplings $J$ but the process weights in the respective phases differ. For the different parameter regimes, the non-vanishing bounce weights are depicted. For large spin-photon coupling the algorithm can be made bounce free.}
	\label{fig:algo_phase_diag}
\end{figure}

One possible solution that maximizes the weights $E,F,G$ and $H$, subsequently maximizes $a, c$ and $d$ and only puts weights to bounce processes if necessary, is given in \Tab{tab:worm-solution}.
\begin{table}[h]
	\begin{center}
\begin{tabular}{ c||c|c|c|c }
	Cases: & {$\Wod > \frac{z}{2} ( \wup + \wAF )$} & \multicolumn{3}{|c}{$\Wod < \frac{z}{2} ( \wup + \wAF )$} \\
 \hline
	$\Bod$ & 0 & \multicolumn{3}{|c}{0}\\
	$D$ & $\Wod - \frac{z}{2}(\wup + \wAF)$ & \multicolumn{3}{|c}{0}\\
 \hline
	& & $\Wod <\frac{z}{2}(\wAF - \wup)$ & $\Wod > \frac{z}{2}\abs{\wAF - \wup}$ &  $\Wod < \frac{z}{2}(\wup - \wAF)$\\
 \hline
	$a$ & 0 & \wup & $\frac{1}{2}(\wup + \wAF) - \frac{\Wod}{z}$ & \wAF \\
	\rmi{b}{\bullet\bullet} & 0 & 0 & 0 & $\wup - \wAF - \frac{2\Wod}{z}$\\
	\rmi{b}{\circ\bullet} & 0 & $\wAF - \wup - \frac{2\Wod}{z}$ & 0 & 0 \\
	$E$ & $\frac{\wup}{2}$ & 0 & $\frac{1}{4}(\wup - \wAF) + \frac{\Wod}{2z}$ & $\frac{\Wod}{z}$ \\
	$F$ & $\frac{\wAF}{2}$ & $\frac{\Wod}{z}$ &  $\frac{1}{4}(\wAF - \wup) + \frac{\Wod}{2z}$ & 0 \\
\hline
	& & $\Wod < \frac{z}{2}(\wAF - \wdown)$ & $\Wod > \frac{z}{2}\abs{\wAF - \wdown}$ &  $\Wod < \frac{z}{2}(\wdown - \wAF)$\\
 \hline
	$c$ & $\frac{1}{2}(\wdown - \wup)$ & \wdown & $\frac{1}{2}(\wdown + \wAF) - \frac{\Wod}{z}$ & \wAF \\
	\rmi{b}{\bullet\circ} & 0 & $\wAF - \wdown - \frac{2\Wod}{z}$  & 0 & 0 \\
	$\rmi{b}{\circ\circ}$ & 0 & 0 & 0 & $\wdown - \wAF - \frac{2\Wod}{z}$\\
	$G$ & $\frac{\wAF}{2} - \frac{1}{4}(\wdown - \wup)$ & $\frac{\Wod}{z}$  & $\frac{1}{4}(\wAF - \wdown) + \frac{\Wod}{2z}$ & 0 \\
	$H$ & $\frac{1}{4}(\wup + \wdown)$ & 0 & $\frac{1}{4}(\wdown - \wAF) - \frac{\Wod}{2z}$ & $\frac{\Wod}{z}$\\
 \hline
\end{tabular}
	\caption{Weights of the loop-construction processes that solve the directed-loop equations as given in Eq.~\ref{eq:worm-equation}. The solution is chosen such that it maximizes the conversion between off-diagonal and diagonal operators and minimizes bounce processes.}
	\label{tab:worm-solution}
\end{center}
\end{table}
The expressions for the process weights depend on how the weights of the vertices compare to each other and one can identify different regions in parameter space.
These different regions can be visualized in an algorithmic phase diagram shown in \Fig{fig:algo_phase_diag}.
For all "algorithmic phases", the non-vanishing bounce weights are shown.
The loop head never bounces at an off-diagonal operator, \ie, $\Bod=0$ for all parameters.

The directed-loop update is started by drawing a random site $\in \{1,\dots, N\}$ and random imaginary time $\in [0, \beta)$.
The loop then starts to propagate in positive or negative direction in imaginary time (with probability $1/2$ respectively) until it encounters a vertex.
The loop is then continued according to the process weights for the vertex encountered.

The actual probabilities that the head of the loop scatters, bounces or straightly traverses a particular vertex are given by the process weights divided by the weight of the vertex. For example, in the case of a diagonal vertex with upward spins, which has weight $\wup$, the probability for the loop head to bounce is $\rmi{p}{up,b} = \bup/\wup$ and the probability to traverse is $\rmi{p}{up,t}=a/\wup$. 
The total probability to have a loop head scatter from this diagonal up-vertex and transforming it into an off-diagonal operator is $\rmi{p}{up,s} = 2E/\rmi{w}{up}$. 
The precise off-diagonal operator is then drawn subsequently. 
The spin site of the exit leg is drawn uniformly from all sites, the imaginary-time displacement $\Delta\tau$ is drawn from the symmetrized photon propagator $\rmi{P}{sym}$ \cite{Weber2022bs} and the subsequent direction of the loop head in imaginary time is drawn with probability $1/2$ (due to the choice $e=e'$).
Instead of drawing directly from the symmetrized photon propagator, one can also draw $\Delta\tau$ from the free photon propagator $P(\omega,\Delta\tau)$ via \cite{Weber2022bs}
\begin{equation}
	\Delta \tau = -\frac{1}{\omega} \ln\left(1 - \xi(1-e^{-\beta\omega})\right)
\end{equation}
with the uniformly distributed random number $\xi \in[0,1)$ and chooses $\tau'= \tau \pm \Delta \tau$ with probability $1/2$ respectively.

For a loop head entering an off-diagonal operator, its bounce probability is $\rmi{p}{od,b} = \Bod/\Wod = 0$ and the probability to traverse is $\rmi{p}{od,t} = D/\Wod$. 
For the more complex scattering process into a diagonal operator, the total probability is $\rmi{p}{od,s} = z(E+F)/\Wod = z(G+H)/\Wod$. 
The precise diagonal operator is then drawn subsequently. 
If the off-diagonal operator connected the space-time coordinates $(i,\tau)$ and $(k,\tau + \Delta\tau)$ and was entered at the latter coordinate, the new diagonal operator connects site $i$ and a neighboring site $j\in \mathrm{nn}(i)$ at imaginary time $\tau$. 
The neighboring site $j$ is drawn uniformly from the nearest neighbors $\mathrm{nn}(i)$ independent of their spin configuration at time $\tau$ (due to the imposed condition $E+F = G+H$).
However, the exit direction of the loop head at site $i$ depends on the spin state at the sites $i$ and $j$. 
If $j$ is $\bullet$, the probability to exit the vertex at site $i$ on the side at which $i$ is $\circ$ such that the new diagonal vertex is $\Diagopd$, is given by $\rmi{p}{\bullet\bullet}=E/(E+F)$. 
The probability for the loop to leave site $i$ into the other direction is consequently given by $\rmi{p}{\circ\bullet}=F/(E+F)$.
Analogously, if $j$ is $\circ$, the probabilities are given by the relative weights of the processes $G$ and $H$ to construct the diagonal vertices $\Diagopaf$ and $\Diagopu$ respectively.
The probabilities for the different processes of a loop entering one of the other three vertex types $\Diagopu, \Diagopaf,\Diagopafr$ can be calculated analogously.

After exiting the updated vertex, the loop continues to propagate in imaginary time until it arrives at the next vertex. 
This procedure is repeated until the loop reaches its starting point and the loop is closed.
The spins along the path of the loop are flipped and another loop can be started.

\subsection{Data structures employed}
In this section we will focus on the employed data structures that differ from the original wormhole algorithm.
Other aspects like the implementation of the ordered vertex list have been adopted from the original algorithm \cite{Weber2022bs} and will not be explained in detail here.

In contrast to other wormhole algorithms employed so far, the wormhole update in our work can also change the sites an operator acts upon. 
This is a consequence of the conversion of off-diagonal operators that couple all pairs of spins to diagonal operators which only couple neighboring spins and vice versa.
Due to this algorithmic difference, the data structures employed for this implementation differ from the ones in the original wormhole algorithms \cite{Weber2017s, Weber2022bs}.
This makes the construction of the loop more dynamic as the scattering processes of the loop are not restricted to specific sites or imaginary times. An operator can potentially be updated several times per wormhole update and wander around unrestrictedly in whole space and imaginary time.
However, this makes it necessary to know the spin state at any point in space and imaginary time during the construction of the loop. 
For this we need a data structure storing the times at which off-diagonal operators are applied to each site.
Such a data structure is also needed in the original wormhole algorithms during the diagonal update \cite{Weber2022bs}. 
However, in the original case one can construct a sorted list of off-diagonal operators per site once before performing the diagonal update. 
During the diagonal update, this list does not need to be updated.
In contrast, during the loop construction of the algorithm in this work, the data structure storing the imaginary times at which an off-diagonal operator is applied needs to be updated frequently. 
For this we chose to employ order statistic trees, which are binary search trees that additionally support the calculation of the index of an element, \ie, how many elements in the tree are smaller than a certain element, in logarithmic time $\mathcal{O}(\ln n)$ in the number of elements $n$.
This tree can be handled separately for ever site such that one has $N$ tree instances with a number of elements $n \sim \beta$ which gives the overall algorithm an additional factor of $\mathcal{O}(\log \beta)$ in its computational complexity with respect to the basic complexity of $\mathcal{O}(\beta N)$.

Another data structure employed in the original wormhole algorithm and also the standard directed-loop algorithm in SSE, is the doubly-linked vertex list. 
This list connects the legs of the vertices in imaginary time and is used for the routing during the loop construction. 
After a vertex is processed and the loop exits a certain leg, the list is used to look up to which leg this exit leg is connected to in imaginary time. 
Again, due to the difference in the present wormhole algorithm of dynamically changing the sites and imaginary times a vertex is related to, this data structure would constantly needed to be updated to insert and remove the legs of altered vertices in between other links.
This is impractical as for the insertion of such links, one would need to lookup which vertex comes before and after a certain imaginary time on a site, which is basically the answer to the question that the linked-vertex list is actually supposed to answer.
So instead, we employed an ordered map for ever site $i$ that maps the imaginary time of a vertex to its index in the vertex list such that one can search for the previous and following vertex in imaginary time on a certain site $i$ considering a certain time $\tau$.
This again gives a factor of $\mathcal{O}(\log \beta)$ with respect to the basic complexity of $\mathcal{O}(\beta N)$.
Overall, the algorithm therefore has a computational complexity of $\mathcal{O}(\beta N \log(\beta))$.
The additional scaling of $\log(\beta)$ in the computational complexity did not have a large effect during our simulations and we managed to simulate a considerable amount of spins (up to $N=8192$, oftentimes $N=2048$ was already sufficient for our purpose) while being sufficiently converged in temperature to sample ground-state properties.
However, when we profiled our code, the insertion and removal of elements in these sorted data structures was nevertheless consuming a considerable amount of the total computation time.
So it might be worthwhile to think about other ways to implement the algorithm, even if it does not realize the best computational complexity achievable.

\subsection{Data analysis tools}
\label{sec:data_analysis}
In the Dicke-Ising model, there exist first-order as well as second-order quantum phase transitions which we have treated differently in terms of data analysis and as a side effect also in terms of data production.

\begin{figure}[h]
	\includegraphics{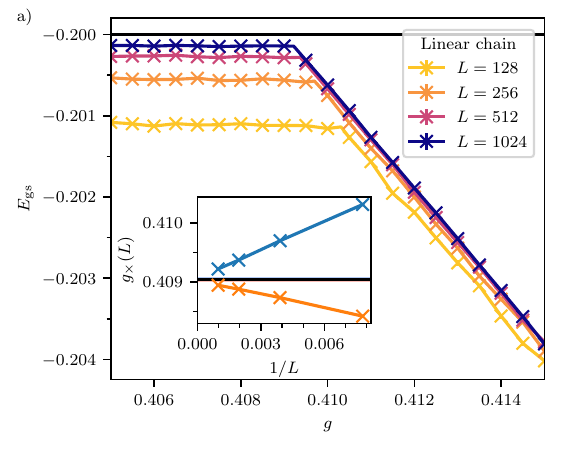}
	\includegraphics{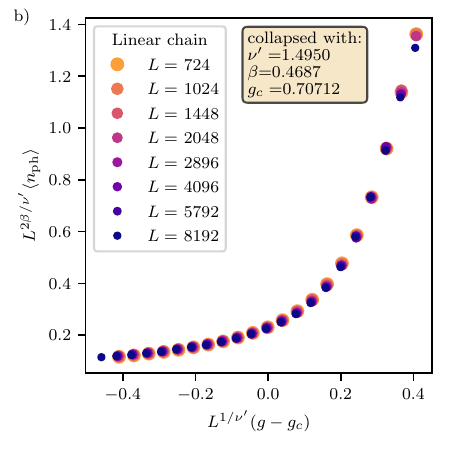}
	\caption{a) Level crossing in the ground-state energy density for the linear chain with vanishing longitudinal field $\epsilon=0$ and $\abs{J}=0.2$. The solid black line is the exact energy density for the AN phase in the thermodynamic limit. Inset: The transition points determined from the kink in the ground state energy density (blue) and the crossing of the energy density in the PS phase on the right with the exact energy density of the AN phase in the thermodynamic limit (orange). The extrapolated transition points are in excellent agreement with the exact value (horizontal black line) calculated in Sec.~\ref{app:sc-mf}. b) Data collapse of the ferromagnetic linear chain with Ising coupling $J=-0.2$ and longitudinal field $\epsilon = 0.6$. Mean field considerations predict the critical point to be at $g_c=1/\sqrt{2}\approx 0.70711$.
	}
	\label{fig:data-analysis}
\end{figure}
For first-order transitions the system exhibits hysteresis across the phase transition. 
We therefore generate data by employing a swipe mechanism to get the energies of the ground states of both phases respectively.
This means we start simulations from both sides of the phase transition and tune the tuning parameter ($g$ or $\epsilon$ depending on the transition) across the transition point ($g_{\cross}$ or $\epsilon_{\cross}$).
An in-depth discussion of this swipe mechanism can be found in Ref.~\cite{Koziol2025s}.

Using this swiped data, we track the level crossings in the ground-state energy density for different system sizes and extrapolate the positions of the level crossing towards the thermodynamic limit $L\rightarrow \infty$. 
The crossing points are calculated by fitting linear functions to the energy densities in the two phases and calculating the crossings of the two fitted functions.
In addition, if one of the phases is a non-superradiant phase, for which the energy density in the thermodynamic limit is known, we also calculate the crossing points of this energy density in the thermodynamic limit with the fitted linear function of the finite-size energy density in the other phase for every $L$.

This procedure is illustrated for vanishing longitudinal field $\epsilon = 0$, Ising coupling $\abs{J}=0.2$ and boson frequency $\omega = 1$ in \Fig{fig:data-analysis} and compared with the exact value from the self-consistent mean field approach in \Sec{app:sc-mf} to demonstrate the high accuracy of the method.
The fitted values for the point of the phase transition are 
\begin{equation}
	\begin{split}
	\rmi{g}{\times,L} &= 0.409062(9)\\
	\rmi{g}{\times,tdl} &= 0.409019(8)
	\end{split}
\end{equation}
for the two methods using the finite-site energy densities in both phases ($\rmi{g}{\times,L}$) and using the energy density in the thermodynamic limit in the non-superradiant phase ($\rmi{g}{\times,tdl}$).
In comparison, the exact value of $\rmi{g}{\times}\approx 0.409044$ from the self-consistent mean field ansatz in \Sec{app:sc-mf} lies in between these two fitted values.

For the second-order quantum phase transitions we perform data collapses of the squared order parameters to determine the critical point and critical exponents $\beta$ and $\nu$. 
The simulation parameters were chosen to lie approximately symmetrically around the critical point and the parameter range was adapted to the expected scaling window.
An exemplary data collapse at the phase transition from normal to superradiant phase for ferromagnetic Ising interaction $J=-0.2$, longitudinal field $\epsilon = 0.6$ and boson energy $\omega=1$ is also shown in \Fig{fig:data-analysis}.
The "critical exponent" $\nu'$ is related to the real critical exponent by $\nu=\koppa\nu'/d=\nu'/\duc$ as explained in the main part.

\section{Classification of quantum phase transitions}
One of the biggest challenges in the analysis of the datasets is their heterogeneity, in particular in terms of the order of the underlying phase transition. The distinction becomes especially difficult close to the multicritical points, where a line of phase transitions changes from first order to second order.
In \Sec{sec:data_analysis} we focused on the conceptual points of the data analysis for first- and second-order quantum phase transitions respectively. 
In this section we want to focus on the physical aspects, \eg, order or criticality of the transition, deduced from the raw data using this previously introduced data analysis tools. 
At the same time we visualize selected datasets behind our results in the main paper regarding the classification of phase transitions and occurrence or absence of multicritical points. 
By this we want to decrease the barrier for the interested reader to sight the raw data, which is also available for individual inspection in Ref.~\cite{RawDatas}.  
We will focus more on the antiferromagnetic Ising case as it is the more complex one, in particular regarding the diversity of phase transitions. Most often we simply note if a certain behavior is also present in the ferromagnetic case.

\subsection{Order of phase transition and multicritical points}
As explained in \Sec{sec:data_analysis}, the point of a first-order transition can be determined by the position of the kink/level crossing in the ground-state energy, while for a second-order transition, one can extract the critical point alongside some critical exponents by performing a data collapse, \eg, of the order parameter.

When approaching the multicritical point on a line of first-order transitions, the level crossing gradually flattens until the kink in the ground-state energy finally disappears at the multicritical point.
This increasingly flat level crossing close to the multicritical point is additionally washed out for finite systems and the flat kink can look more like a bending of the energy for systems sizes not large enough.
The ground-state energy for small systems can therefore look continuous well ahead of the multicritical point, which makes the correct classification in this regime extremely challenging and limited by the maximum system size we can achieve within the simulations.

\begin{figure}
	\includegraphics{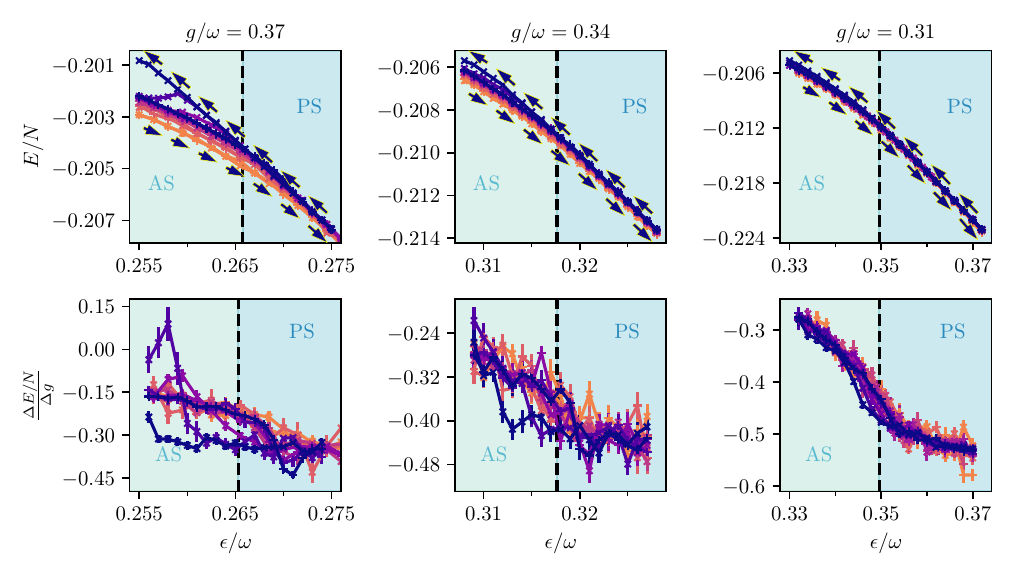}
	\caption{Softening of the level crossing close to the multicritical point for the transition between antiferromagnetic superradiant (AS) and paramagnetic superradiant (PS) phase on the square lattice for $g/\omega=0.37,0.34,0.31$ (from left to right). The upper row shows the energy density while the lower row shows the numerical derivative of this energy density. A kink in the energy density corresponds to a jump in the derivative of the energy density. The multicritical point was estimated to be $g\approx 0.30(2)$ by additionally taking into account the breakdown of the data collapse coming from the opposite side of the multicritical point. The simulated system sizes are $L=12,16,22,32,46,64$ and additionally $L=90$ for $g/\omega=0.31$ closest to the multicritical point. The color ranges from bright orange for the smallest to dark blue for the largest system size. 
	For the energy density, the swipe direction from left to right (right to left) is visualized by the arrow $\rightarrow$ ($\leftarrow$) and marker $+$ ($\cross$).
}
	\label{fig:softening}
\end{figure}

The described softening and flattening of the first-order transition occurs for the transition between normal and superradiant phase in the ferromagnetic model on both lattices as well as for the magnetic transition between the intermediate antiferromagnetic superradiant and paramagnetic superradiant phase for the antiferromagnetic model on the square lattice.
For the last case, the gradual softening is shown in \Fig{fig:softening} as a representative example by showing the energy density and its numerical derivative for $g/\omega=0.37,0.34,0.31$.
The multicritical point was estimated to be at $g\approx 0.30(2)$ in this case.
While for $g/\omega = 0.37$, the kink is still clearly visible for the largest simulated system sizes, for $g/\omega=0.34$ one can only see a tiny hysteresis for the largest system size and for $g/\omega = 0.31$ the ground-state energy looks continuous for all simulated system sizes (up to $N=90\times90=8100$). 
However, when looking at the numerical derivative of the energy density (lower row in \Fig{fig:softening}), one can still see a tiny hysteresis and discontinuity even for $g/\omega=0.31$, indicating a first-order transition.

\begin{figure}[h!]
	\includegraphics{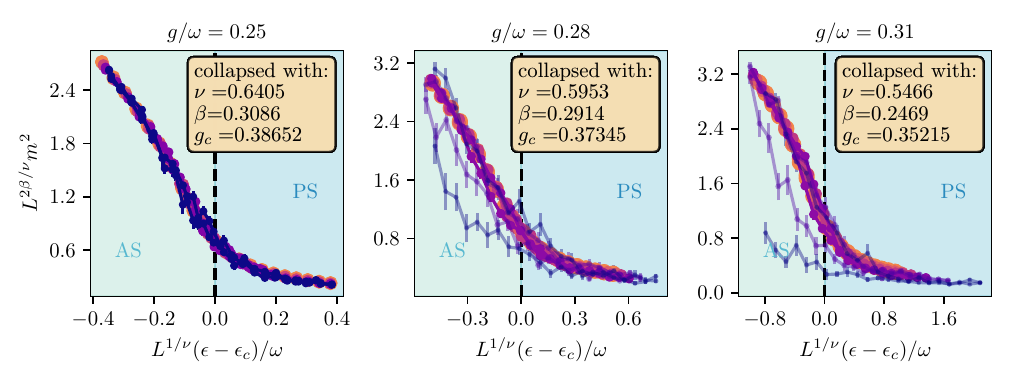}
	\caption{Data collapse close to the multicritical point for the transition between antiferromagnetic superradiant and paramagnetic superradiant phase (AS-PS) on the antiferromagnetic square lattice for $g/\omega=0.25,0.28,0.31$ (from left to right). The multicritical point was estimated to be $g\approx 0.30(2)$ by additionally taking into account the softening of the kink in the ground-state energy density coming from the other side of the multicritical point. The transparent data points for large system sizes were not used in the data collapse due to their hysteresis. The simulated system sizes are $L=12,16,22,32$ and additionally $L=46,64$ for $g/\omega = 0.28,0.31$. The color ranges from bright orange for the smallest to dark blue for the largest system size. }
	\label{fig:worsening_dc}
\end{figure}
\begin{figure}[h!]
	\includegraphics{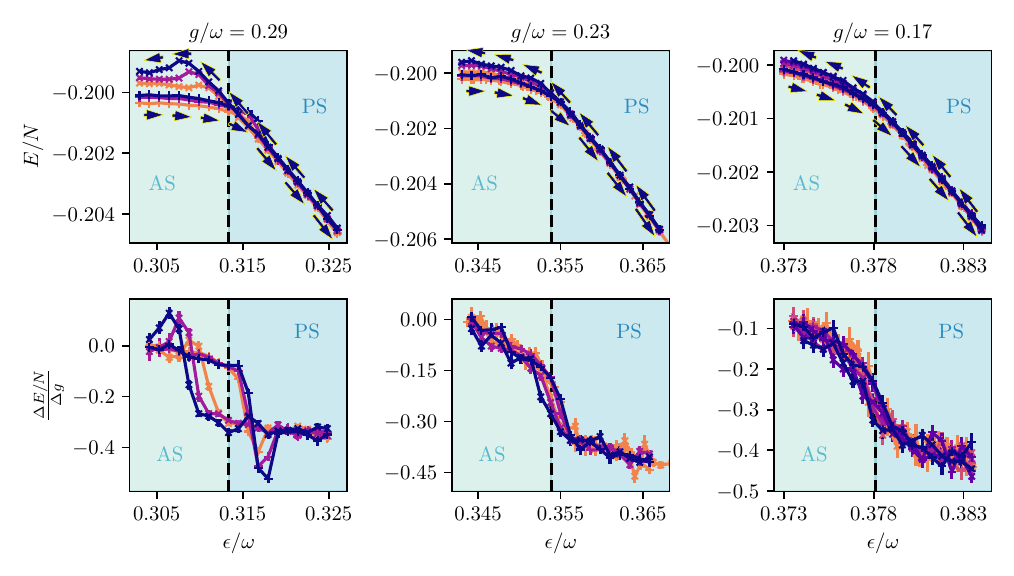}
	\caption{Energy density (upper row) and its numerical derivative (lower row) between antiferromagnetic superradiant (AS) and paramagnetic superradiant (PS) phase on the chain for $g/\omega=0.29, 0.23, 0.17$ showing a gradual softening of the level crossing. The softening is also visible in the derivative of the energy density by the softening of the hysteresis/jump. All datasets still show a level crossing. The simulated system sizes are $L=256, 512, 1024$ and additionally $L=2048$ for $g/\omega=0.17$. The color ranges from bright orange for the smallest to dark blue for the largest system size.}
	\label{fig:softening_1d}
\end{figure}

On the opposite side of a multicritical point, the data is analyzed by performing a data collapse of the respective order parameter. In addition to the critical point, one additionally obtains the critical exponents $\beta$ and $\nu$ from this.
For transitions between normal and superradiant phase, the order parameter is the photon density and the extracted critical exponents belong to the same universality class as the Dicke model ($\beta = 1/2$ and $\nu^\prime = 3/2$ as discussed in the main body of the paper).
For transitions between paramagnetic and antiferromagnetic phase, the order parameter is the staggered squared magnetization and the extracted critical exponents belong to the 3d Ising universality class \cite{Kos2016s} as discussed in the main part of the paper.

When approaching the multicritical point, the order parameter starts to display a hysteresis behaviour for larger system sizes and one needs to exclude these from the analysis to get a consistent data collapse. 
In addition, the fitted exponents start to shift away from the respective universality class and the collapsed curves for different $L$ start to be sheared against each other.
This worsening of the data collapse and the hysteresis behaviour is shown in \Fig{fig:worsening_dc}.

In the case of the antiferromagnetic chain, the simulated datasets showed no indication of a second-order magnetic phase transition between the antiferromagnetic superradiant and paramagnetic superradiant phase in contradiction to a mean field analysis \cite{Zhang2014s}.
However there is still a softening of the level crossing taking place for decreasing values of $g/\omega$, which is shown in \Fig{fig:softening_1d}.

\subsection{Critical properties}
As there are several second-order phase transitions in the phase diagrams of the Dicke-Ising model, one natural question is to which universality class these transitions belong to.
In the main paper we stated that all second-order transitions between normal and superradiant phases have the same criticality as the Dicke model, which is consistent with the mapping of non-superradiant phases to the Dicke model \cite{Schellenberger2024s}.
For the magnetic transition between antiferromagnetic superradiant and paramagnetic superradiant phase on the square lattice for antiferromagnetic Ising interactions we assigned the 3d Ising universality class in the main part of the paper.
As we performed data collapses for every dataset at second-order phase transitions, we do not get a single set of critical exponents but get slightly different critical exponents for each of the datasets.
\begin{figure}[h]
	\includegraphics{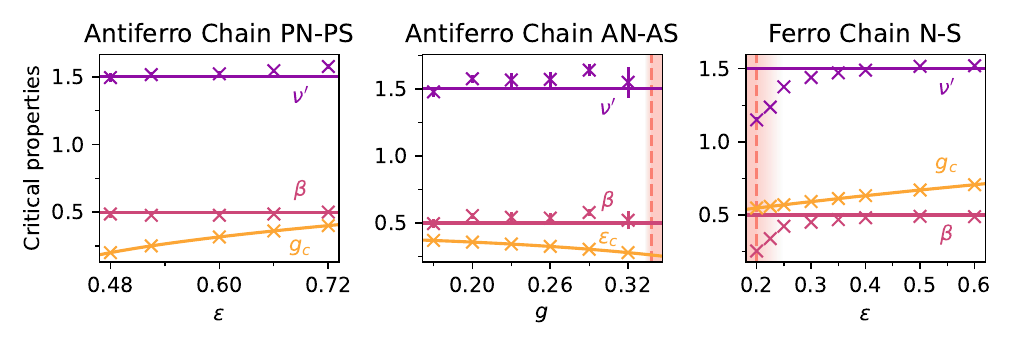}
	\includegraphics{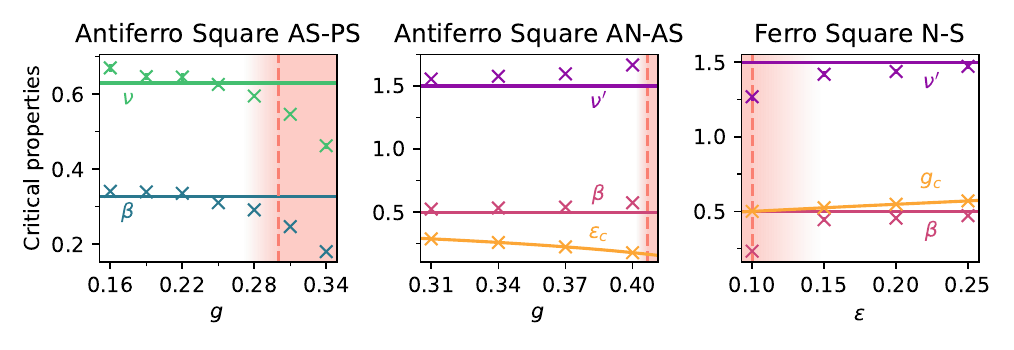}
	\caption{Critical properties extracted by the method of data collapse of the respective squared order parameters are compared to the critical exponents of the respective universality class and, in the case of normal to superradiant transition, the exactly known transition point. The critical exponents between normal and superradiant phases belong to the same universality class as the Dicke model ($\beta=1/2$, $\nu^\prime = 3/2$). The critical exponents of the magnetic transition AS-PS on the square lattice (lower left) belong to the 3d Ising universality class \cite{Kos2016s}.}
	\label{fig:crit_exp}
\end{figure}

In \Fig{fig:crit_exp}, we show the extracted critical exponents for all second-order transitions and compare them to the ones of the respective assigned universality class. The multicritical points are depicted by dashed vertical lines, except for the transition from paramagnetic normal to paramagnetic superradiant phase (PN-PS) as the whole line is of second-order.
The red-shaded region visualizes the region beyond the multicritical point, where the transition is of first order and a data collapse is not an appropriate tool to analyze the phase transition. The blurriness of the red-shaded region gives an estimate for the uncertainty of the multicritical point.
One can see that the exponents shift away from the respective universality class, when entering the shaded region as the data collapse starts to be of poor quality due to the closeness to the multicritical point.

\section{Self-consistent mean field calculation for the quantized transverse-field Ising model}
\label{app:sc-mf}
In the following we present a self-consistent mean-field calculation for the Dicke-Ising model with $\epsilon=0$ which allows to determine the location of the first-order phase transition up to numerical precision for the one-dimensional chain.
The ground state of the magnetically-ordered normal phase is a product state of zero bosons and the classical Ising ground state \cite{Rohn2020s,Schellenberger2024s}. 
Its ground-state energy density is therefore $e = E/N = -\abs{J}$. In the following we determine the ground-state energy density of the paramagnetic superrradiant phase.

 With 
\begin{equation*}
	S_x = \sum_{i=1}^{N} \sigma_i^x/2
\end{equation*}
\begin{comment}
\begin{align}
	\Ham =  \omega a^\dagger a  +  2\frac{g}{\sqrt{N}} (a^\dagger + a)S_x + J \sum_{\lla i, j\rra} \sz_i \sz_j
	\label{eq:dicke-ising}
\end{align}
\end{comment}
we introduce a conditional displacement transformation 
\begin{equation}
	U = \exp\left(\frac{2g}{\sqrt{N}} \,S_x\,(a^{\phantom\dagger}-a^\dagger)\right)
\end{equation}
in the same way as in \cite{Rohn2020s}. 
In the superradiant phase $U$ transforms the ground state of $\Ham$ to a state with zero photon density. 
In the displaced basis, the breakdown of the superradiant phase is signaled by a finite photon density.
The Hamiltonian $\Ham = \mathcal{H}_{\mathrm{Dicke}} + \mathcal{H}_{\mathrm{Ising}}$ in this new basis is
\begin{equation}
	\begin{aligned}
		U^\dagger \mathcal{H} U = \omega a^\dagger a - 4 g^2 S_x^2/N + U^\dagger \mathcal{H}_{\mathrm{Ising}} U.
	\end{aligned}
	\label{eq::transformedHam_notexpanded}
\end{equation}
We will show that the ground-state energy density of this Hamiltonian is equal to that of 
\begin{equation}
	\omega a^\dagger a - 4 g^2 S_x^2/N + \mathcal{H}_{\mathrm{Ising}}
\end{equation}
in the superradiant phase. 
One has
\begin{equation}
	\begin{aligned}
	U^\dagger \mathcal{H}_{\mathrm{Ising}} U & = \mathcal{H}_{\mathrm{Ising}} +
	\frac{2g}{\sqrt{N}} \,(a^{\phantom\dagger}-a^\dagger) 	 \left[\mathcal{H}_{\mathrm{Ising}},  \,S_x \right] + 1/2 
	\left(\frac{2g}{\sqrt{N}} (a^{\phantom\dagger}-a^\dagger)\right)^2
	\left[\left[\mathcal{H}_{\mathrm{Ising}}, S_x \right], S_x \right] + \dots \\
	& = \mathcal{H}_{\mathrm{Ising}} + R.
    \end{aligned}
    \label{eq::transformedHamIsing_expanded}
\end{equation}

The commutator $[\mathcal{H}_{\mathrm{Ising}},S_x]$ yields an extensive number of local operators. 
The same holds for all higher-order commutators in Eq.~\eqref{eq::transformedHamIsing_expanded} which can be seen iteratively. 
As a consequence, because $\Braket{a}/\sqrt{N} = 0$, the terms in $R$ scale sub-extensively with $N$. 
Hence, $R/N$ vanishes in the thermodynamic limit and the ground-state energy density of the superradiant phase in the thermodynamic limit is the same as for $\omega a^\dagger a - 4 g^2 S_x^2/N + \mathcal{H}_{\mathrm{Ising}}$.

One can simplify the decoupled matter part further by rewriting
\begin{equation*}
	S_x = \Braket{S_x} + \delta S_x \equiv N \cdot m + \delta S_x 
\end{equation*}
and assuming that the fluctuations are subextensive:
\begin{equation}
	S_x^2/N = N m^2 + 2 m \, \delta S_x + \delta S_x^2 /N \approx N m^2 + 2 m \,\delta S_x = - N m^2 + 2 m S_x \, .
\end{equation}
Plugging this into Eq.~\eqref{eq::transformedHam_notexpanded} yields
\begin{equation}
	U^\dagger \mathcal{H} U = \omega a^\dagger a + J \sum_{\lla i, j\rra} \sz_i \sz_j +4g^2 N m^2 - 8g^2 m S_x \equiv \tilde{H}(g,m)
\end{equation}
and the self-consistent equation
\begin{equation}
	\Braket{S_x}_{\tilde{H}(g,m)}/N = m \,.
\end{equation}
The matter part corresponds therefore to a self-consistent transverse-field Ising model (TFIM). 
In a similar fashion a broad class of light-matter Hamiltonians was reduced to a self-consistent matter problem in \cite{roman2024linears,roman2024cavitys} including finite temperatures.

In one dimension, the TFIM is exactly solvable which allows to determine the location of the phase transition exactly (see also \cite{roman2024linears,roman2024cavitys}). 
In the thermodynamic limit we have \cite{pfeuty1970ones}
\begin{equation}
	m = \left(L(0) + \lambda L(1)\right)/2
\end{equation}
with 
\begin{equation}
	L(n) = \frac{1}{\pi} \int_{0}^{\pi}\, dk \Lambda_k^{-1} \cos(kn),
\end{equation}
\begin{equation}
	\Lambda_k^2 = 1+\lambda^2+2\lambda\cos(k),
\end{equation}
and
\begin{equation}
	\lambda = \frac{J}{2g^2 m}.
\end{equation}
Now $m$ and the ground-state energy density 
\begin{equation}
	E/N \equiv e  = - 4 g^2 m/ \pi \int_{0}^{\pi}\, dk \Lambda_k + 4 g^2 m^2
\end{equation}
can be obtained up to machine precision numerically.
A clear first-order transition is found because this energy density crosses the energy density $-\abs{J}$ of the magnetically-ordered normal phase before the magnetization in the self-consistent loop goes to zero. The phase transition is located at $g_c = 0.9146494478 \sqrt{J}$.

\section{Effects of a diamagnetic $A^2$-term on the phase diagrams}
\label{app:A-squared}
In this section we briefly discuss the effect of including a diamagnetic $A^2$-term 
in the Dicke-Ising Hamiltonian
\begin{align}
	\Ham &=  \omega a^\dagger a  + \epsilon \sum_{i=1}^N{\sz_i} + \frac{g}{\sqrt{N}} (a^\dagger + a) \sum_{i=1}^N \sigma^x_i + J \sum_{\lla i, j\rra} \sz_i \sz_j + \Ham_{A^2}\\
	\Ham_{A^2} &= D (a^\dagger + a)^2\, ,
\end{align}
which commonly arises from the minimal-coupling substitution $p \rightarrow p - qA$ when deriving the Dicke model microscopically \cite{Rzazewski1975s, FriskKockum2019s, Garziano2020s}.

In the Dicke model ($J=0$), this $A^2$-term is one origin of no-go theorems prohibiting the superradiant phase transition \cite{Rzazewski1975s} as it suppresses the formation of a coherent photon state with finite photon density.
Including this term therefore can lead to qualitatively different behavior and makes it necessary to take this term into account, in particular when claiming the formation of a superradiant phase.
However, there is no benefit in explicitly including the $A^2$-term in our simulations as this term can be absorbed into the photon energy by a Bogoliubov transformation which renormalizes the photon frequency $\omega \rightarrow \omega'$ \cite{FriskKockum2019s} and spin-photon coupling $g \rightarrow g'$
\begin{equation}
	\begin{split}
		\omega' &= \sqrt{\omega^2 + 4\omega D}\\
		g' &= \sqrt{\frac{\omega}{\omega'}} g.
	\end{split}
	\label{eq:renormalized_photon_frequency}
\end{equation}
The quantum phase diagrams calculated for the pure Dicke-Ising model without the diamagnetic $A^2$-term already include the full information for the Dicke-Ising model with the diamagnetic term.
When including the diamagnetic term, the phase diagrams will be distorted and the accessible regions in the phase diagrams will be limited by the renormalization of the photon frequency \Eq{eq:renormalized_photon_frequency}.

The coefficient $D$ of the diamagnetic term is in general not a free parameter.
In certain realizations of the Dicke model including the $A^2$-term \cite{Rzazewski1975s, Viehmann2011s}, the Thomas-Reiche-Kuhn (TRK) sum rule gives a lower bound $D\geq g^2/2\epsilon$ with respect to the other parameters.
We decided to take the lower bound $D=g^2/2\epsilon$ of the TRK sum rule for the Dicke-Ising model for illustrating its effect on the phase diagrams. 
In principle, one can also use other values for $D$ and transform the phase diagram post simulation.

\begin{figure}
	\includegraphics{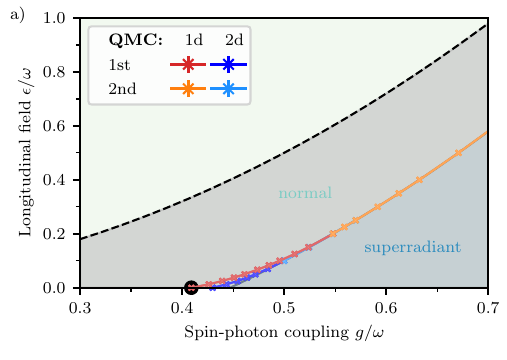}
	\includegraphics{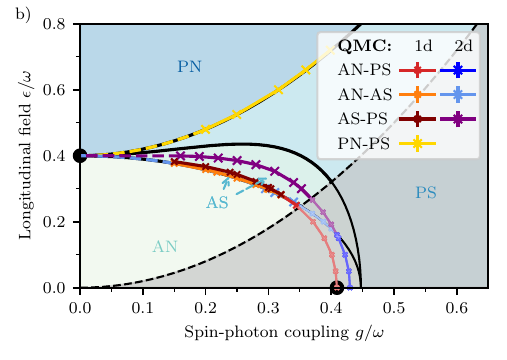}
	\caption{Quantum phase diagrams from main paper showing, in addition, the boundary of accessible parameter space (dashed black lines). The grey-shaded region below the black dashed lines cannot be reached when including an $A^2$-term respecting the TRK sum rule giving a lower bound on the coefficient $D\geq g^2/2\epsilon$ of the $A^2$-term. In the limiting case of the Dicke model, the boundary of the inaccessible parameter space coincides with the transition line from normal to superradiant phase. a) In the ferromagnetic case, the superradiant phase can not be realized. b) In the antiferromagnetic phase diagram, one can still reach all four phases including the superradiant ones for finite $\epsilon$ and $J$.}
	\label{fig:phase_diagram_a2}
\end{figure}

In \Fig{fig:phase_diagram_a2}, we show the same quantum phase diagrams for ferromagnetic and antiferromagnetic Ising interaction as in the main part but with an additional boundary that separates the accessible and inaccessible regions of the phase diagrams in the presence of a diamagnetic term with $D=g^2/2\epsilon$. 
The distortion is not shown as we want to focus on the restrictions of the phase diagrams in comparison to the case without diamagnetic term.
In the case of the ferromagnetic Ising interactions, the superradiant phase completely lies within the inaccessible region of the phase diagram.
However, in the antiferromagnetic case, in which the costs for local spin excitations in the normal phase are lowered with respect to the Dicke model, the superradiant phases, including the intermediate AS phase, also extends to the accessible region of the phase diagram.
In the Dicke case ($J = 0$), the boundary would lie on top of the transition line from normal to superradiant phase as expected for the borderline case of $D=g^2/2\epsilon$.

\end{document}